\documentclass[11pt, a4paper]{article}
\pdfoutput=1
\usepackage[T1]{fontenc}
\usepackage[utf8]{inputenc}
\usepackage[english]{babel}
\usepackage{jcappub}
\usepackage{natbib}
\usepackage{amsmath}
\usepackage{amssymb}
\usepackage{graphicx}
\usepackage{xcolor, color}
\usepackage{multirow}
\usepackage{tensor}
\usepackage{siunitx}
\usepackage{soul}

\usepackage{cleveref}

\usepackage{url}
\usepackage{xspace}
\usepackage{subcaption}
\usepackage{bm}
\usepackage{braket}
\usepackage{amssymb}
\usepackage{amsmath}
\newcommand*\Bell{\ensuremath{\boldsymbol\ell}}
\newcommand{\mvec}[1]{\bm{#1}}

\newcommand{\myvec}[1]{\mvec{#1}}

\graphicspath{{./Immagini/}}

\title{Breaking degeneracies with the Sunyaev-Zeldovich full bispectrum}

\author[a,1]{Andrea Ravenni,\note{Corresponding author.}}
\author[b,c,d]{Matteo Rizzato,}
\author[e]{Slađana Radinović,}
\author[f,g,h]{Michele Liguori,}
\author[i]{Fabien Lacasa,}
\author[d]{and Elena Sellentin}

\affiliation[a]{Jodrell Bank Centre for Astrophysics, School of Physics and Astronomy, The University of Manchester, Manchester M13 9PL, U.K.}
\affiliation[b]{Sorbonne Universit\'e, CNRS, UMR 7095, Institut d'Astrophysique de Paris, 98 bis bd Arago, 75014 Paris, France}
\affiliation[c]{Sorbonne Universit\'es, Institut Lagrange de Paris (ILP), 98 bis bd Arago, 75014 Paris, France}
\affiliation[d]{Leiden Observatory, Leiden University, Huygens Laboratory, Niels Bohrweg 2, NL-2333 CA Leiden, The Netherlands}
\affiliation[e]{Institute of Theoretical Astrophysics, University of Oslo, 0315 Oslo, Norway}
\affiliation[f]{Dipartimento di Fisica e Astronomia “G. Galilei”, Università degli Studi di Padova, via~Marzolo~8, I-35131, Padova, Italy}
\affiliation[g]{INFN, Sezione di Padova, via~Marzolo~8, I-35131, Padova, Italy}
\affiliation[h]{INAF-Osservatorio Astronomico di Padova, vicolo~dell'Osservatorio~5, I-35122 Padova, Italy}
\affiliation[i]{Université Paris-Saclay, CNRS, Institut d’astrophysique spatiale, 91405, Orsay, France}
\emailAdd{andrea.ravenni@manchester.ac.uk}
\emailAdd{rizzato@strw.leidenuniv.nl}
\emailAdd{sladana.radinovic@astro.uio.no}
\emailAdd{liguori@pd.infn.it}
\emailAdd{fabien.lacasa@universite-paris-saclay.fr}
\emailAdd{sellentin@strw.leidenuniv.nl}

\newcommand{\diff}{\mathop{}\!\mathrm{d}}
\newcommand\Diff[1]{\mathop{}\!\mathrm{d^#1}}
\newcommand{\td}[2]{\frac{\diff #1}{\diff #2}}

\newcommand{\vers}[1]{\hat{#1}}

\newcommand{\Planck}{\textit{Planck}\xspace}

\newcommand{\Thomson}{Thomson\xspace}

\newcommand{\sigmaT}{\sigma_\text{T}}

\newcommand{\Te}{T_\text{e}}
\newcommand{\me}{m_\text{e}}

\abstract{Non-Gaussian (NG) statistics of the thermal Sunyaev-Zeldovich (tSZ) effect carry significant information which is not contained in the power spectrum. Here, we perform a joint Fisher analysis of the tSZ power spectrum and bispectrum to verify how much the full bispectrum can contribute to improve parameter constraints. We go beyond similar studies of this kind in several respects: first of all, we include the complete power spectrum and bispectrum 
(auto- and cross-) covariance in the analysis, computing all NG contributions; furthermore we consider a multi-component foreground scenario and model the effects of component separation in the forecasts; finally, we consider an extended set of both cosmological and intra-cluster medium parameters. We show that the tSZ bispectrum is very efficient at breaking parameter degeneracies, making it able to produce even stronger cosmological constraints than the tSZ power spectrum: e.g. the standard deviation on $\sigma_8$ shrinks from $\sigma^\text{PS}(\sigma_8)=0.35$ to $\sigma^\text{BS}(\sigma_8)=0.065$ when we consider a multi-parameter analysis. We find that this is mostly due to the different response of separate triangle types (e.g. equilateral and squeezed) to changes in model parameters. While weak, this shape dependence is clearly non-negligible for cosmological parameters, and it is even stronger, as expected, for intra-cluster medium parameters.}

\keywords{Sunyaev-Zeldovich effect --- cosmological parameters from LSS}

\begin{document}
\maketitle
\flushbottom
\newpage

\section{Introduction}

The thermal Sunyaev-Zeldovich (tSZ) effect \cite{Zeldovich:1969ff, Sunyaev:1970er} \cite[for a recent review]{Mroczkowski:2018nrv} is a spectral distortion of the Cosmic Microwave Background (CMB), mostly generated in galaxy clusters by inverse Compton scattering of CMB photons off hot electrons.
The tSZ effect is a powerful cosmological observable, mainly applied to the study of individual clusters, to build cluster catalogues and to extract number count statistics. A complementary possibility consists in the study of the tSZ angular power spectrum. After being originally discussed in \cite{Komatsu:2002wc}, this approach has then been adopted as a powerful probe of the low-redshift Universe, to test both the standard $\Lambda$CDM scenario \cite{Ade:2013qta, Aghanim:2015eva} and some extended models, which encompass primordial non-Gaussianity, massive neutrinos and dark energy \cite{Hill:2013baa, Roncarelli:2014jla, Bolliet:2017lha, McCarthy:2017csu, Bolliet:2019zuz}.
One of the advantages of the tSZ power spectrum analysis is that it allows including also small, unresolved clusters, and it does not require direct measurements of cluster masses.

As it has been argued long before the tSZ was routinely measured across the sky \cite{Cooray:2000xh}, it is important to notice that the tSZ map is highly non-Gaussian, therefore only a part of the available tSZ information is actually captured by the power spectrum. A natural question which arises is therefore how much additional information can be extracted from higher order statistics, starting with the bispectrum (i.e. the 3-point multipole correlation function).

An initial theoretical study of the tSZ bispectrum was performed in \cite{2012ApJ...760....5B}. There, it was pointed out that, besides having a different amplitude scaling with $\sigma_8$, compared to the power spectrum, the bispectrum signal also takes its main contributions from massive clusters at low redshift. This makes the impact of astrophysical uncertainties smaller in the bispectrum than in the power spectrum, since the latter has sizeable contribution from less-well-understood, low-mass, high-redshift clusters. 
At the same time the tSZ skewness was detected using Atacama Cosmology Telescope (ACT) data \cite{2012PhRvD..86l2005W, Hill:2012ec}, while an all-sky Compton-$y$ map, and subsequent measurements of both the skewness and the bispectrum were later obtained by {\it Planck} \cite{Ade:2013qta, Aghanim:2015eva}. In both cases it was shown that the bispectrum could be used to obtain significant constraints on $\sigma_8$. A later study \cite{Hurier:2017jgi} combined cluster counts with power spectrum and equilateral (all three $\ell$'s equal) bispectrum measurements, showing that the bispectrum --- even just by considering the small equilateral triangle subset --- can play a significant role in breaking (cosmological and astrophysical) parameter degeneracies.
If, as we will show, this is the case, the tSZ bispectrum would be an extremely valuable source of information, especially in light of how difficult it is proving to understand the astrophysics of halos and reconcile various measurement of gas parameters, particularly the hydro-static mass bias $b_\text{HSM}$
\cite{Zubeldia:2019brr,Makiya:2018pda, Makiya:2019lvm,Salvati:2019zdp,Ruppin:2019deo,Pandey:2019uxy}.

These and other similar results clearly encourage further investigation. The aim of this work is to explore how much extra-information is contained in the 3-point function, via a detailed joint Fisher analysis of the tSZ power spectrum and bispectrum. Our main goal is to make our forecasts as realistic and accurate as possible. For this purpose:
\begin{itemize}
\item We consider the entire bispectrum domain (not just specific triangles) and compute for the first time the full tSZ bispectrum covariance, beyond the Gaussian approximation. We therefore include all contributions in the bispectrum covariance, up to the connected 6-point correlator, and we will also account for correlations between the power spectrum and the bispectrum.
\item We extend our parameter space with respect to previous forecasts and analyses, in order to model in greater detail the impact of uncertainties in the electron pressure profile.
\item We account for foreground contamination and model (in a simple way) the effects of component separation, using Internal Linear Combination (ILC).
\end{itemize}
As we show in the following, our analysis reinforces the conclusion that the tSZ bispectrum is a powerful observable to constrain cosmology, especially due to its efficiency in breaking degeneracies between astrophysical and cosmological parameters, otherwise present in a power spectrum-only analysis. 

As a note to our previous statements, let us also stress here that the bispectrum does not account for all the cosmological information which can be extracted from the NG component of the tSZ map, since of course relevant contributions also come from higher order correlators.
As long as most of the information is captured by the amplitude --- and not by the shape --- of different tSZ correlation functions, it was shown that the tSZ 1-point probability-density-function is a near-optimal statistic to constrain cosmology, since it encodes information from all $n$-point amplitudes (see \cite{Thiele:2018jdl} and references therein).
The implementation of the 1-point statistic in parameter inference is however work in progress at the moment, making bispectrum estimation still a viable and worth pursuing alternative approach in practice. Moreover, we will show in the remainder that, while indeed small, the tSZ bispectrum shape dependence on cosmology is not negligible, and it actually plays an important role in breaking parameter degeneracies; this argument becomes even stronger when we account for the shape dependence on IntraCluster Medium (ICM) parameters.

The paper is structured as follows: in section \ref{sec:theorycorr} we review the halo model approach applied to the theoretical calculation of tSZ $n$-point correlation function, we show our choice of halo mass function and bias and we describe the analytical model of electron pressure profile which we use in the analysis; in section \ref{3.2} we illustrate the calculation of the full covariance matrix for our observables; in section \ref{sec:Derivatives} we analyze the dependence of the tSZ power spectrum and bispectrum on different cosmological and ICM parameters; in section \ref{sec:GaussianForegrounds} we discuss foreground contamination issues; in section \ref{sec:ForecastResults} we describe in detail and comment the results of our analysis. Finally, we summarize and draw our conclusions in section \ref{sec:Conclusions}.

\section{Theoretical tSZ correlation functions}\label{sec:theorycorr}
The temperature fluctuations associated with the tSZ effect appear in sky maps as
\begin{equation}
    \frac{\Delta T}{T}(\nu,{\hat{n}})
    =
    g(\nu) \, y({\hat{n}}) \, ,
\quad
    g(\nu)
    \equiv
    x \coth (x/2) - 4 \, ,
\quad
	y({\hat{n}})
	\equiv
	\int \sigmaT \frac{k_\text{B} \, n_\text{e} \Te}{\me c^2}
	\diff r \, .
\label{eq:y-distortion}
\end{equation}
Here $g(\nu)$ encapsulates the spectral dependence of the tSZ effect in units of CMB temperature with $x \equiv h \nu/(k_\text{B} T)$, 
while $y$, called the Compton-$y$ parameter, is the tSZ intensity along the line of sight $\vers{n}$. $\sigmaT$ is the \Thomson cross section and $m_\text{e}$, $n_\text{e}$ and $\Te$ are respectively the electron rest mass, number density and temperature.

The statistical properties of the $y$ field can be extracted from its $n$-point angular correlation functions that we predict using the halo model approach \cite[for a review]{Cooray:2002dia}.
Under some simplifying assumptions, discussed later, the expressions for all $n$-point correlators are well known \cite{Komatsu:2002wc, Aghanim:2015eva, Lacasa:2014gha, 2012ApJ...760....5B, Hurier:2017jgi, Salvati:2017rsn}.
They can be derived via the formalism detailed in appendix~\ref{sec:A_ProjectedAngularCorrelators}, where we express the $y$-polyspectra as the projection onto the past light-cone of the three-dimensional correlators of the corresponding three-dimensional field. This formulation is very useful in the mathematical derivation and numerical implementation of the general $y$ field $n$-point correlator.

The tSZ power spectrum is given by the sum of the Poisson one-halo term, which account for intra-halo correlations, and the two-halo term which models inter-halo correlations \cite{Hill:2013baa,2011MNRAS.418.2207T}.
They read
\begin{gather}
C_\ell^{1\mathrm{h}}  = \int \diff z \frac{\Diff{2} V}{\diff z \diff \Omega} \int \diff M \frac{\diff n}{\diff M} (z, M) |\tilde{y}_\ell(z,\tilde{M})|^2 \, ,
\label{eq:Clyy1h}
\\
C_\ell^{2\mathrm{h}}  = \int \diff z \frac{\Diff{2} V}{\diff z \diff \Omega} D_+^2(z) P_m(k) \bigg[\int \diff M \frac{\diff n}{\diff M} (z, M)  b(z, M) \tilde{y}_\ell(z,\tilde{M}) \bigg]^2\bigg|_{k=\big(\frac{\ell + 1/2}{\chi(z)}\big)} .
\label{eq:Clyy2h}
\end{gather}
Here $P_m(k)$ is the linear matter power spectrum, $D_+(z)$ is the growth factor, and $\Diff{2} V/ \diff z \diff \Omega$ is the comoving volume element per steradian, which can be calculated as $\Diff{2} V/ \diff z \diff \Omega = c\chi^2(z)/H(z)$, where $\chi(z)$ is the radial comoving distance.
The Halo Mass Function (HMF) $\frac{\diff n}{\diff M}$ and halo bias $b$ will be described in section \ref{sec:HMF}.
$\tilde{y}_\ell(z,M)$ is the 2D Fourier transform of the projected $y$ parameter image of the halo defined in detail in section \ref{sec:ElectronPressureProfile}.
Notice that in eq. \eqref{eq:Clyy1h}, \eqref{eq:Clyy2h}, and throughout the paper
we also consider a hydrostatic mass bias rescaling the true mass of the halo, i.e., we use $\tilde{M} = (1-b_\text{HSM})M$ in the expression for the projected~$y$. 

Higher order correlators are a straightforward extension of the power spectrum.
Throughout this work, we will adopt the flat-sky approximation.
Therefore, we will have to include the reduced bispectrum in our NG analysis. The bispectrum is described by one-, two- and three-halo terms.
Including the three halo term would require the second order halo bias, which is not provided in \cite{2010ApJ...724..878T}.
To overcome this shortcoming, one could in principle resort to the peak-background-split formalism \citep{1984ApJ...284L...9K,1986ApJ...304...15B,1989MNRAS.237.1127C,1996MNRAS.282..347M};
however, since the three-halo contribution is negligible we just omit it here.
The one- and two-halo terms read
\begin{align}
b_{\ell_1 \ell_2 \ell_3}^{1\mathrm{h}}  
	= 
	\, &
	\int \diff z 
	\frac{\Diff{2} V}{\diff z \diff \Omega} 
	\int \diff M \frac{\diff n}{\diff M} (z, M) 
	\tilde{y}_{\ell_1}(z,\tilde{M})
	\tilde{y}_{\ell_2}(z,\tilde{M})
	\tilde{y}_{\ell_3}(z,\tilde{M})
	\, ,
\label{eq:Byyy1h}
\\
\begin{split}
	b_{\ell_1 \ell_2 \ell_3}^{2\mathrm{h}}  
	=
	\, &
	\Bigg[
	\int \diff z \frac{\Diff{2} V}{\diff z \diff \Omega} 
	D_+^2(z) P_m(k) 
	\int \diff M 
	\frac{\diff n}{\diff M} (z, M)  b(z, M) 
	\tilde{y}_{\ell_2}(z,\tilde{M}) \, \tilde{y}_{\ell_3}(z,\tilde{M}) \times 
\\
	\, &
	\times 
	\int \diff M 
	\frac{\diff n}{\diff M} (z, M)  b(z, M) 
	\tilde{y}_{\ell_1}(z,\tilde{M})\bigg|_{k=\big(\frac{\ell_1 + 1/2}{\chi(z)}\big)}
	\Bigg]
	+ \text{2 permutations.}
\end{split}
\label{eq:Byyy2h}
\end{align}
Finally, for the higher order correlators, we will always employ only the one-halo term, which in general reads
\begin{equation}
\label{General1halo}
P_{(n)}^{1\mathrm{h}}  (\ell_1, \dots, \ell_n)
	= 
	\int \diff z 
	\frac{\Diff{2} V}{\diff z \diff \Omega} 
	\int \diff M \frac{\diff n}{\diff M} (z, M) 
	\prod_i
	\tilde{y}_{\ell_i}(z,\tilde{M}) \,
	.
\end{equation}
For later use, we point out that the general one-halo term does depend solely on the magnitudes of the multipoles involved.

Unless specified otherwise, in all the spectra we integrate over redshift between $z_\text{min}=10^{-6}$ and $z_\text{max}=4.5$, and over masses between $M_\text{min}=10^{10}\, M_\odot h^{-1}$ and $M_\text{max}=10^{16}\, M_\odot h^{-1}$.
These limits, that stretch beyond the values commonly used in the literature, allow us to ensure that even higher order correlators are integrated correctly.
We integrate directly over the overdensity mass $M_{500,c}$, as in, e.g., \cite{Aghanim:2015eva, Bolliet:2017lha}, so that a conversion to the virial masses is never required, and also assume that $\diff \ln M_{\Delta,c} / \diff \ln M_\text{vir} \approx 1$ \cite{Bolliet:2017lha}.
We use $b_\text{HSM}=0.2$ \cite{Aghanim:2015eva} as the fiducial value throughout the analysis.
We choose the values from \cite{Ade:2015xua} for the cosmological parameters $h= 0.6711$, $n_S= 0.9624$, $w_0= -1$, but we pick $\Omega_m= 0.28$, $\sigma_8= 0.8$ to be in agreement with \cite{Aghanim:2015eva}.%
\footnote{For consistency with \cite{Aghanim:2015eva}, we do not use the updated parameter from \cite{Aghanim:2018eyx}.}

The fitting functions we employ, that we are about to describe, are the most commonly used in the literature. A possible alternative, proposed in \cite{Mead:2020qgo}, would be to jointly fit both the pressure profile and the HMF from simulations

\subsection{Halo mass function}
\label{sec:HMF}
We use the HMF and bias from \cite{2010ApJ...724..878T}, converting the mass definition $M_{500,c}$ into $M_{\Delta,m}$, to fit the parameters of their table 4 to the appropriate value of $\Delta = 500 \rho_c / \rho_m$.

\begin{equation}
    \td{n}{M}
    =
    \nu f(\nu)
    \frac{\bar{\rho}_m}{M}
    \td{\ln(\sigma^{-1})}{M} \, ,
\qquad
    f(\nu) 
    =
    \alpha \left[
        1+(\beta \nu)^{-2\phi}
    \right]
    \nu^{2\eta}
    e^{-\gamma \nu^2/2} \, .
\end{equation}
We assume that the redshift scaling provided for $\Delta=200$ applies to any $\Delta$:
\begin{equation}
    \beta= \beta_0 (1+z)^{0.2} \, , \quad
    \phi= \phi_0 (1+z)^{-0.08} \, , \quad
    \eta= \eta_0 (1+z)^{0.27} \, , \quad
    \gamma= \gamma_0 (1+z)^{-0.01} \, ,
\end{equation}
where the parameters at $z=0$ are taken from table 4 of \cite{2010ApJ...724..878T}, using a linear interpolation along $\Delta$.
As recommended, for each $z>3$ we use the value calculated at $z=3$, and at each $z$ we calculate $\alpha(z)$ imposing
$
\int \diff \nu f(\nu,z) \, b(\nu,z) = 1 \, .
$
At $z=0$ we correctly recover the value of $\alpha$ interpolated from the above mentioned table.

We use a halo bias of the form
\begin{equation}
    b(\nu) = 1- A\frac{\nu^a}{\nu^a + \delta_c^a} + B\nu^b + C\nu^c \, ,
\end{equation}{}
with the parameters from table 2 of \cite{2010ApJ...724..878T}.

\subsection{Electron pressure profile}
\label{sec:ElectronPressureProfile}
In our analysis, we are interested in the projected two-dimensional Compton-$y$ field, which is obtained via line-of-sight integration of the rescaled electron density profile $P_{\mathrm{e}}$, in any given direction of the sky through
\begin{equation}
    \label{2Dcomptony}
	\tilde{y}_\ell(z,M) 
	=
	\frac{\sigmaT}{\me c^2}
	\frac{4\pi r_{s}}{\ell_{s}^2} 
	\int \diff x x^2 \,
	j_0\!\left(\frac{\ell + 1/2}{\ell_s \chi(x)} x\right) P_{\mathrm{e}}(x,z,M) \, .
\end{equation}
Here $r_{s} = r_{s}(z,M)$ and $\ell_{s} = a(z) \chi(z)/r_{s}(z,M)$ are, respectively, the typical scale radius of the $y$-image of the halo, and the multipole moment associated with it.
For the parametrization of the electron profile that we employ, given in \cite{Arnaud:2009tt}, $r_s = r_{500}$ and, for a single cluster, $P_{\mathrm{e}}$ relies on the generalised Navarro-Frenk-White profile
\begin{equation}
\label{Pe}
    P_{\mathrm{e}}\left(x, z, M_{500,c}\right) = \frac{C\left(x, z, M_{500,c}\right)  \times P_0}{\left(c_{500}x\right)^{\gamma_G}\left[1+\left(c_{500}x\right)^{\alpha_G}\right]^{\frac{\gamma_G-\beta_G}{\alpha_G}}}\,,\quad x \equiv \frac{r}{r_{500}} \, ,
\end{equation}
for which we use the best fit parameters obtained in \cite{Arnaud:2009tt}:%
\footnote{We use the non-self-similar profile derived in the main text (cf. their eq. (12)).}%
\begin{equation}
\label{Arnaud}
    \{P_0, c_{500}, \alpha_G,\gamma_G,\beta_G\} = \big \{ 8.403 \ h_{70}^{-3/2}, 1.177, 0.3081, 1.0510, 5.4905 \big \} \, .
\end{equation}
As we rely on the halo model for the description of matter clustering, $r_{500}$ is the distance from the center of the halo, delimiting a sphere containing a density of dark matter which is $500$ times the critical density of the Universe $\rho_{\mathrm{c}}$.
The function $C\left(x,z,M_{500,c}\right)$ has the expression 
\begin{equation}
\label{prefactorC}
    C\left(x,z,M_{500,c}\right) = 1.65\ h_{70}^2\  E(z)^{\frac{8}{3}} 
    \left(
        \frac{
            h_{70}
            \,  M_{500,c}
        }{3\times 10^{14}\  M_{\odot}}\right)^{\frac{3}{2} + \alpha_{\mathrm{P}} + \alpha'(x)}\ \mathrm{eV}\ \mathrm{cm}^{-3},
\end{equation}
where $\alpha_{\mathrm{P}} = 0.12$, and the exponent $\alpha'_\mathrm{P}(x)$, also fitted in \cite{Arnaud:2009tt}, has the following dependence 
\begin{equation}
\alpha_\mathrm{P}'(x) = 0.10 - \left(\alpha_{\mathrm{P}}+0.10\right)\frac{\left(\frac{x}{0.5}\right)^3}{1+\left(\frac{x}{0.5}\right)^3} \, .
\end{equation}
These exponents parametrise deviation from the standard self-similar case. In eqs. \eqref{Arnaud}-\eqref{prefactorC} we also introduced the widely used parameter $h_{70} \equiv h/0.7$.

\section{Covariance matrix for the (binned) observables}
\label{3.2}

In this work, we consider flat-sky binned estimators for the observables of interest as they are smooth and slowly varying functions in multipole space. The bins are defined in harmonic space: $\Bell' \in \ell^b$ if $\ell - \Delta\ell^b/2 \le \ell' \le \ell + \Delta\ell^b/2$, $\ell$ being the central value of the bin $\ell^b$ of width $\Delta\ell^b$. As we are working in flat-sky, $\Bell\in\mathbb{R}^2$. We are also considering a survey observing a sky fraction corresponding to a solid angle of $\Omega_{\mathrm{sky}}$ steradians. Similarly to the three-dimensional matter field case, the binned angular power spectrum is defined as \citep{2006MNRAS.371.1188H,2007NJPh....9..446T,2012JCAP...04..019D}
\begin{equation}
\label{binnedCl}
    \hat{C}_{\ell^b}
    \equiv
    \sum_{\Bell \in \ell^b} 
    \frac{\delta\left(\Bell+\Bell'\right)\ \tilde{y}_{\Bell}\ \tilde{y}_{\Bell'}}{\Omega_{\mathrm{sky}}N(\ell^b)} \, ,
\end{equation}
where the sum runs over discrete modes which are integer multiples of the fundamental frequency of our survey $\ell_{\mathrm{f}} \equiv 2\pi/\Theta_{\text{sky}}$, $\Theta_{\text{sky}}$ being the survey linear angular size. We recall that
\begin{equation}
\Omega_{\text{sky}} = 2\pi\left(1-\cos\Theta_{\text{sky}}\right).    
\end{equation}
The field $\tilde{y}_{\Bell}$ is the two-dimensional Fourier transform of the $y$ field \eqref{eq:y-distortion}. The quantity $N(\ell^{\mathrm{b}})\approx 2\ell\Delta\ell^{\mathrm{b}} f_{\text{sky}}$ (in the limit $\ell\gg\ell_f$) gives the number of  vector pairs $\hat{\Bell},-\hat{\Bell}$ whose magnitude $\hat{\ell}$ is within the bin $\ell^{\mathrm{b}}$, each pair being discriminated by a deviation in the module of the vectors of a unit of the survey fundamental mode $\ell_{\mathrm{f}}$ \citep{2008A&A...477...43J,2009A&A...508.1193J,2013MNRAS.429..344K}.
In a similar fashion to the three-dimensional matter field case, we define the binned angular bispectrum as \cite{1998ApJ...496..586S,2004PhRvD..69j3513S,2013MNRAS.429..344K,Bucher:2015ura,2018PhRvD..97d3532C}
\begin{equation}
\label{binnedbll}
    \hat{b}_{\ell^b_1 \ell^b_2 \ell^b_3}
    \equiv
    \sum_{\Bell_1 \in \ell_1^b}
    \sum_{\Bell_2 \in \ell_2^b}
    \sum_{\Bell_3 \in \ell_3^b}
    \frac{ \delta^{(2)}(\Bell_1 + \Bell_2 + \Bell_3)\ \tilde{y}_{\Bell_1}\tilde{y}_{\Bell_2}\tilde{y}_{\Bell_3}}{\Omega_{\mathrm{sky}}N_{\mathrm{tri.}}(\ell_1^b, \ell_2^b, \ell_3^b)} \, ,
\end{equation}
where $N_{\mathrm{tri.}}(\ell_1^b, \ell_2^b, \ell_3^b)$ is the number of valid triplets (i.e. respecting the triangular delta) in the proposed bin triplets \citep{2008A&A...477...43J,2009A&A...508.1193J,2013MNRAS.429..344K}
\begin{equation}
N_{\text{tri.}}\left(\ell_1^{\mathrm{b}}, \ell_2^{\mathrm{b}}, \ell_3^{\mathrm{b}}\right) \approx \frac{\Omega^2_{\mathrm{sky}}\ \ell_1\ \ell_2\ \ell_3\ \Delta\ell_1^{\mathrm{b}}\ \Delta\ell_2^{\mathrm{b}}\ \Delta\ell_3^{\mathrm{b}}}{2\pi^3\sqrt{2\ell_1^2\ell_2^2 + 2\ell_1^2\ell_3^2 + 2\ell_2^2\ell_3^2 - \ell_1^4 - \ell_2^4 - \ell_3^2}} \, ,
\end{equation} 
where the approximation assumed is $\Delta\ell_i^{\mathrm{b}} \gg \ell_f$. 
To speed up the calculations of the Fisher matrix, we assume that the power spectrum and the bispectrum are constant in each bin $\ell^{\mathrm{b}}$, and can therefore be described by a single suitably picked representative $\bar{\ell}^{\mathrm{b}}$.%
\footnote{We discuss and validate our choice of binning and bin representative in appendix \ref{A_Binning}.}
In practice, it is possible to prove that the 2 estimators above are unbiased and  $\langle\hat{C}_{\ell^b}\rangle = C_{\bar{\ell}^b}$ and $\langle\hat{b}_{\ell^b_1 \ell^b_2 \ell^b_3}\rangle = b_{\bar{\ell}^b_1 \bar{\ell}^b_2 \bar{\ell}^b_3}$.
In the following we drop the bars to distinguish the bin from its representative, as they should be distinguishable from the context. The representative considered in this work will be the central magnitude value $\bar{\ell}^{\mathrm{b}} \equiv \ell$ of the bin $\ell^{\mathrm{b}}$.

\subsection{Structure of the covariance matrix}
For our computation, we chose the binned estimators $\hat{C}_{\ell^{\mathrm{b}}}$ \eqref{binnedCl} and  $\hat{b}_{\ell_1\ell_2\ell_3}$ \eqref{binnedbll} for which we present the covariance matrix here below.
We split the joint covariance as \citep{2018JCAP...06..015B,2018JCAP...10..053B,2019MNRAS.490.4688R,2019arXiv190806234S}
\begin{align}
\text{Cov}\left[\hat{C}_{\ell^{\mathrm{b}}} ,\hat{C}_{\ell^{'\mathrm{b}}}\right] &= \text{Cov}\left[\dots \right]_{\text{Gauss}} + \text{Cov}\left[\dots \right]_{\text{NG}},\label{CovPPsynt}\\
\text{Cov}\left[\hat{b}_{\ell^{\mathrm{b}}_1\ell^{\mathrm{b}}_2\ell^{\mathrm{b}}_3},\hat{b}_{\ell^{'\mathrm{b}}_1\ell^{'\mathrm{b}}_2\ell^{'\mathrm{b}}_3} \right] &=\text{Cov}\left[\dots\right]_{\text{Gauss}} + \text{Cov}\left[\dots\right]_{\text{NG}}, \label{CovBBsynt}\\
\text{Cov}\left[\hat{C}_{\ell^{\mathrm{b}}},\hat{b}_{\ell^{\mathrm{b}}_1\ell^{\mathrm{b}}_2\ell^{\mathrm{b}}_3} \right] &=
\text{Cov}\left[\dots\right]_{\text{NG}}. \label{CovPBsynt}
\end{align}
In eqs.~\eqref{CovPPsynt} and \eqref{CovBBsynt}, the subscript Gauss labels the covariance terms containing only 2-point statistics, which are non-vanishing only for correlations within the same $\Bell-$bin. The other covariance terms arise due to the non-Gaussian statistics of the $y$ field \eqref{eq:y-distortion} and correlate modes in different $\Bell-$bins and the power spectrum and bispectrum.  The derivation of the flat-sky joint power spectrum-bispectrum covariance has already been outlined for other projected scalar fields, such as the weak lensing convergence \citep{2013MNRAS.429..344K,2019MNRAS.490.4688R}. As the derivation of the covariance for the $y$ field follows the same mathematical and conceptual steps, we refer to the above cited works for more insights on its rigorous derivation. In the following, we report the final expressions employed in the present work, along with the most important concepts associated to their structure.

\subsection{Power spectrum covariance matrix}
Starting from the power spectrum binned estimator \eqref{binnedCl}, we can obtain its full covariance by applying the following standard definition 
\begin{equation}
\text{Cov}\left[\hat{C}_{\ell^{\mathrm{b}}} ,\hat{C}_{\ell^{'\mathrm{b}}}\right] = \langle\hat{C}_{\ell^{\mathrm{b}}}\hat{C}_{\ell^{'\mathrm{b}}}\rangle - \langle\hat{C}_{\ell^{\mathrm{b}}}\rangle \langle\hat{C}_{\ell^{\mathrm{b}}}\rangle \, . 
\end{equation} 
A detailed calculation leads to a non-connected correlator of 4 instances of the field $\tilde{y}_{\ell}$ which can be split via Wick theorem into a sum of products of 2-point correlators and one 4-point connected correlator. The former can be further simplified, leading to the following Gaussian term
\begin{equation}
\text{Cov}\left[\hat{C}_{\ell^{\mathrm{b}}} ,\hat{C}_{\ell^{'\mathrm{b}}} \right]_{\text{Gauss}} = \frac{2\delta^{\mathrm{K}}_{\ell\ell'}}{N\left(\ell^{\mathrm{b}}\right)}\left( C_{\ell}^{\mathrm{n.}} \right)^2 ,
\qquad
C^{\mathrm{n.}}_{\ell} \equiv C_{\ell} + N_{\ell} \, .
\label{CovPPG}
\end{equation}
In eq. \eqref{CovPPG} we  potentially account for sources of Gaussian noise via the associated power spectrum $N_{\ell}$.
The 4-point connected component leads instead to the NG term in eq.~\eqref{CovPPsynt}
\begin{equation}
\text{Cov}\left[\hat{C}_{\ell^{\mathrm{b}}} ,\hat{C}_{\ell^{'\mathrm{b}}} \right]_{\text{NG,T}} \approx \frac{1}{\Omega_{\mathrm{sky}}}T\big(\ell,-\ell,\ell',-\ell'\big) \, ,\label{CovPPNG}
\end{equation}
where $T\big(\ell,-\ell,\ell',-\ell'\big)$ is the trispectrum \eqref{RenameT} of the tSZ field.
In eq.~\eqref{CovPPNG} the exact covariance evaluation would require an average of the trispectrum over the two bins $\ell^{\mathrm{b}}, {\ell'}^{\mathrm{b}}$ involving both angular and magnitude integrations. However, we work under the common approximation of slowly varying polyspectra within the bins chosen for our work \citep{2013MNRAS.429..344K,2019MNRAS.490.4688R}. We therefore assumed the averaged trispectrum being the same as the trispectrum computed at the central values of the bins.

\subsection{Bispectrum covariance matrix}

\begin{figure}
    \centering
    \includegraphics[width=\linewidth]{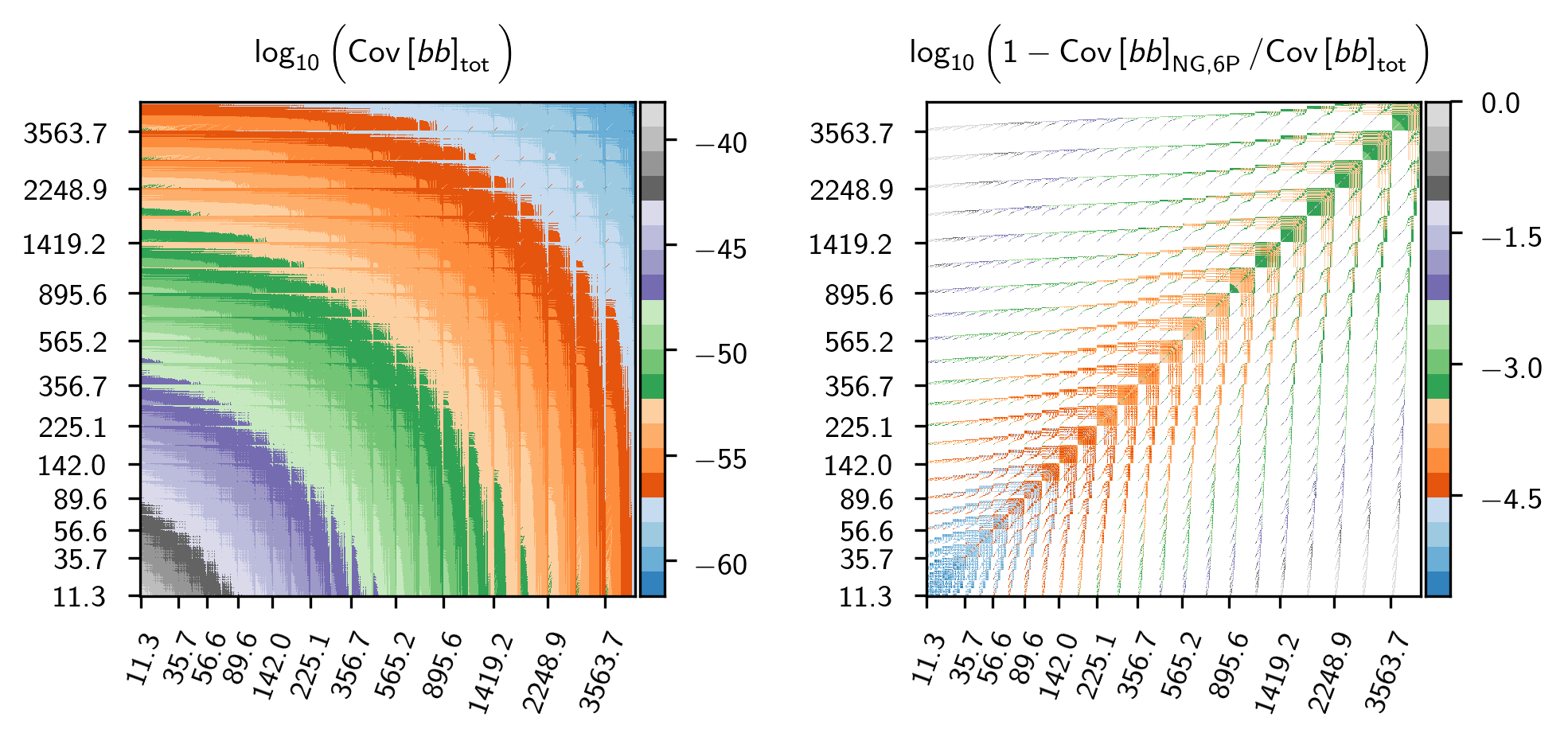}
    \caption{\textit{Left)} Full bispectrum covariance matrix.
    \textit{Right)} Fractional amplitude of the bispectrum covariance terms different from $\text{Cov}\big[\hat{b}_{\ell^{\mathrm{b}}_1\ell^{\mathrm{b}}_2\ell^{\mathrm{b}}_3},\hat{b}_{\ell^{'\mathrm{b}}_1\ell^{'\mathrm{b}}_2\ell^{'\mathrm{b}}_3} \big]_{\text{NG,6P}}$ compared to the full covariance.
    The white areas correspond to configurations for which the covariance is fully determined by $\mathrm{Cov}_{\mathrm{BB}}^{(6)}$ for which the plotted quantity is not defined.
    In both panels, the ticks mark equilateral configurations with sides of the labelled length.}
    \label{fig:CovMat_BS}
\end{figure}

Similarly to the procedure outlined in the previous section, we can compute the covariance matrix for the flat-sky bispectrum binned estimator \eqref{binnedbll}, shown in figure \ref{fig:CovMat_BS}, left panel.
\citep{2013MNRAS.429..344K,2013arXiv1306.4684K,2019JCAP...03..008B}.
The Wick theorem allows us to derive the Gaussian part of the bispectrum covariance
\begin{equation}
    \begin{split}
        \text{Cov}\Big[\hat{b}_{\ell^{\mathrm{b}}_1\ell^{\mathrm{b}}_2\ell^{\mathrm{b}}_3},\hat{b}_{\ell^{'\mathrm{b}}_1\ell^{'\mathrm{b}}_2\ell^{'\mathrm{b}}_3} \Big]_{\text{Gauss}}
        =
        \frac{\Omega_{\text{sky}}}{N_{\text{tri.}}\left(\ell_1^{\mathrm{b}},\ell_2^{\mathrm{b}},\ell_3^{\mathrm{b}}\right)}
        C_{\ell_1}^{\mathrm{n.}}C_{\ell_2}^{\mathrm{n.}}C_{\ell_3}^{\mathrm{n.}}
        \Big[ \delta^{\mathrm{K}}_{\ell_1\ell'_1}\left(\delta^{\mathrm{K}}_{\ell_2\ell'_2}\delta^{\mathrm{K}}_{\ell_3\ell'_3} + \delta^{\mathrm{K}}_{\ell_2\ell'_3}\delta^{\mathrm{K}}_{\ell_3\ell'_2}\right) + 
    \hspace{20pt} &
\\
    + \text{2 permutations of}\ \big(  \ell'_1\leftrightarrow \ell'_2 \big) +
	\text{2 permutations of}\ \big(  \ell'_1\leftrightarrow \ell'_3 \big) \Big].
	&
    \end{split}
\end{equation}
Meanwhile, the non-Gaussian (NG) components read
\begin{equation}
\begin{split}
    \text{Cov}
    \Big[
    	\hat{b}_{\ell^{\mathrm{b}}_1\ell^{\mathrm{b}}_2\ell^{\mathrm{b}}_3},
    	\hat{b}_{\ell^{'\mathrm{b}}_1\ell^{'\mathrm{b}}_2\ell^{'\mathrm{b}}_3} 
    \Big]_{\text{NG,BB}}
    = 
    \frac{2\pi}{\Omega_{\text{sky}}}b_{\ell'_1,\ell_2,\ell_3}b_{\ell_1,\ell'_2,\ell'_3}
    \bigg[ \frac{1}{\ell_1\Delta\ell_1^{\mathrm{b}}}\left(\delta^{\mathrm{K}}_{\ell_1\ell'_1}+\delta^{\mathrm{K}}_{\ell_1\ell'_2}+\delta^{\mathrm{K}}_{\ell_1\ell'_3}\right) +
    \hspace{15pt} &
\\
	+ \text{3 permutations of}\ \big(  \ell_1\leftrightarrow \ell_2 \big) 
	+ \text{3 permutations of}\ \big(  \ell_1\leftrightarrow \ell_3 \big)\bigg],
	&
\\
\end{split}
\label{CovBB}
\end{equation}
\begin{equation}
\begin{split}
    \text{Cov}
    \Big[
    	\hat{b}_{\ell^{\mathrm{b}}_1\ell^{\mathrm{b}}_2\ell^{\mathrm{b}}_3},
    	\hat{b}_{\ell^{'\mathrm{b}}_1\ell^{'\mathrm{b}}_2\ell^{'\mathrm{b}}_3} 
    \Big]_{\text{NG,PT}}
    = 
    \frac{2\pi}{\Omega_{\text{sky}}}
    \frac{C_{\ell_1}^{\mathrm{n.}}}{\ell_1\Delta \ell_1^{\mathrm{b}}}
    \left[
        T\left(\ell_2,\ell_3,\ell'_2,\ell'_3\right)\delta^{\mathrm{K}}_{\ell_1\ell'_1}+
        T\left(\ell_2,\ell_3,\ell'_1,\ell'_3\right)\delta^{\mathrm{K}}_{\ell_1\ell'_2}+
    \right.&
\\
	\left. 
    	+ T\left(\ell_2,\ell_3,\ell'_1,\ell'_2\right)\delta^{\mathrm{K}}_{\ell_1\ell'_3}
    \right] +
    \text{3 permutations of}\ \big(  \ell_1\leftrightarrow \ell_2 \big) +
    \text{3 permutations of}\ \big(  \ell_1\leftrightarrow \ell_3 \big).&
\end{split}
\label{CovPT}
\end{equation}

By employing the estimator definitions, we can also obtain an expression for the cross-covariance
between the binned flat-sky observables used in our analyses
\begin{equation}
\label{Covpb}
\begin{split}
	\text{Cov}\Big[C_{\ell^{\mathrm{b}}},\hat{b}_{\ell^{\mathrm{b}}_1\ell^{\mathrm{b}}_2\ell^{\mathrm{b}}_3}\Big]_{\text{NG,PB}}
    =
    &
    \frac{4\pi}{\Omega_{\text{sky}}}
    \bigg[ 
    \frac{C_{\ell}^{\mathrm{n.}}\left(\ell\right)b_{\ell,\ell_2,\ell_3}}{\ell_1\Delta\ell_1^{\mathrm{b}}}\delta^{\mathrm{K}}_{\ell\ell_1}
    + \text{1 permutation of}\ \big( \ell_1\leftrightarrow \ell_2 \big) + 
\\
    &
	+ \text{1 permutation of}\ \big(  \ell_1\leftrightarrow \ell_3 \big)\bigg].
\end{split}
\end{equation}
So far, all the equations listed in the present section are exact. The two terms still missing are those associated to the 6- and 5-point correlator arising from the bispectrum covariance and the cross-covariance calculation. In these cases, the spectra should be averaged over the bins involved. However, as it was the case for the NG component of the power spectrum covariance \eqref{CovPPNG}, we assume them to be almost constant over the bins thus avoiding this computationally expensive operation. Therefore, the final expressions are
\begin{align}
   \text{Cov}\Big[\hat{b}_{\ell^{\mathrm{b}}_1\ell^{\mathrm{b}}_2\ell^{\mathrm{b}}_3},\hat{b}_{\ell^{'\mathrm{b}}_1\ell^{'\mathrm{b}}_2\ell^{'\mathrm{b}}_3} \Big]_{\text{NG,6P}} 
   &
   \approx  \frac{1}{\Omega_{\mathrm{sky}}}P_{(6)}\left(\ell_1,\ell_2,\ell_3,\ell'_1,\ell'_2,\ell'_3\right), \label{Ins6P}
\\
   \text{Cov}\Big[\hat{C}_{\ell^{\mathrm{b}}},\hat{b}_{\ell^{\mathrm{b}}_1\ell^{\mathrm{b}}_2\ell^{\mathrm{b}}_3}\Big]_{\text{NG,5P}} 
   & 
   \approx \frac{1}{\Omega_{\mathrm{sky}}}P_{(5)} \left(\ell,-\ell,\ell_1,\ell_2,\ell_3\right). \label{Ins5P}
\end{align}

\begin{figure}[t]
    \centering
    \includegraphics[width=\linewidth]{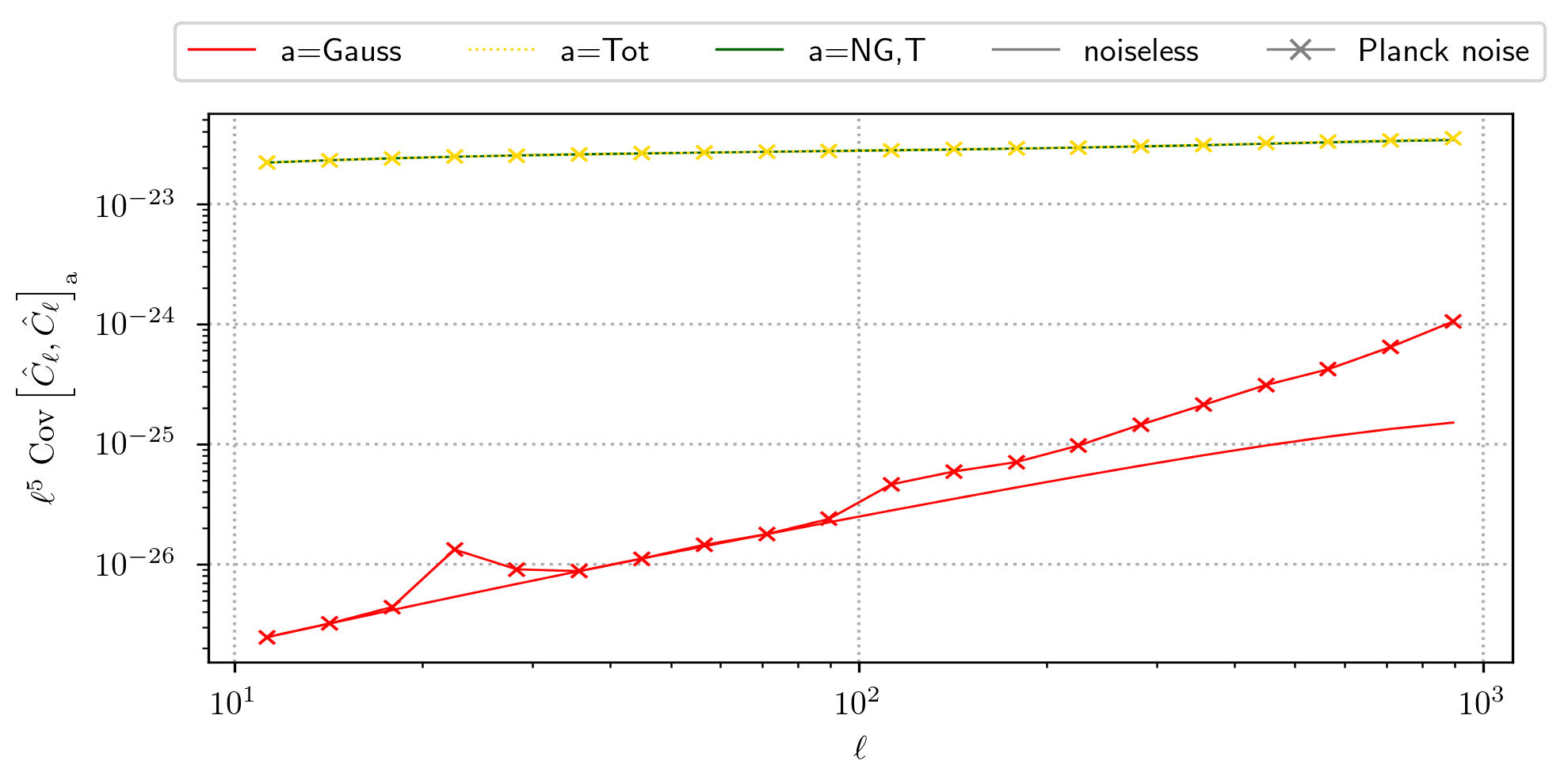}
    \caption{Different components of the diagonal part for the tSZ power spectrum covariance matrix. The impact of the Gaussian Planck-like noise is displayed through makers.
    All the components are properly scaled for visualisation purpose.}
    \label{fig:CovarianceHierarchy_PS}
\end{figure}
\begin{figure}
    \centering
    \includegraphics[width=\linewidth]{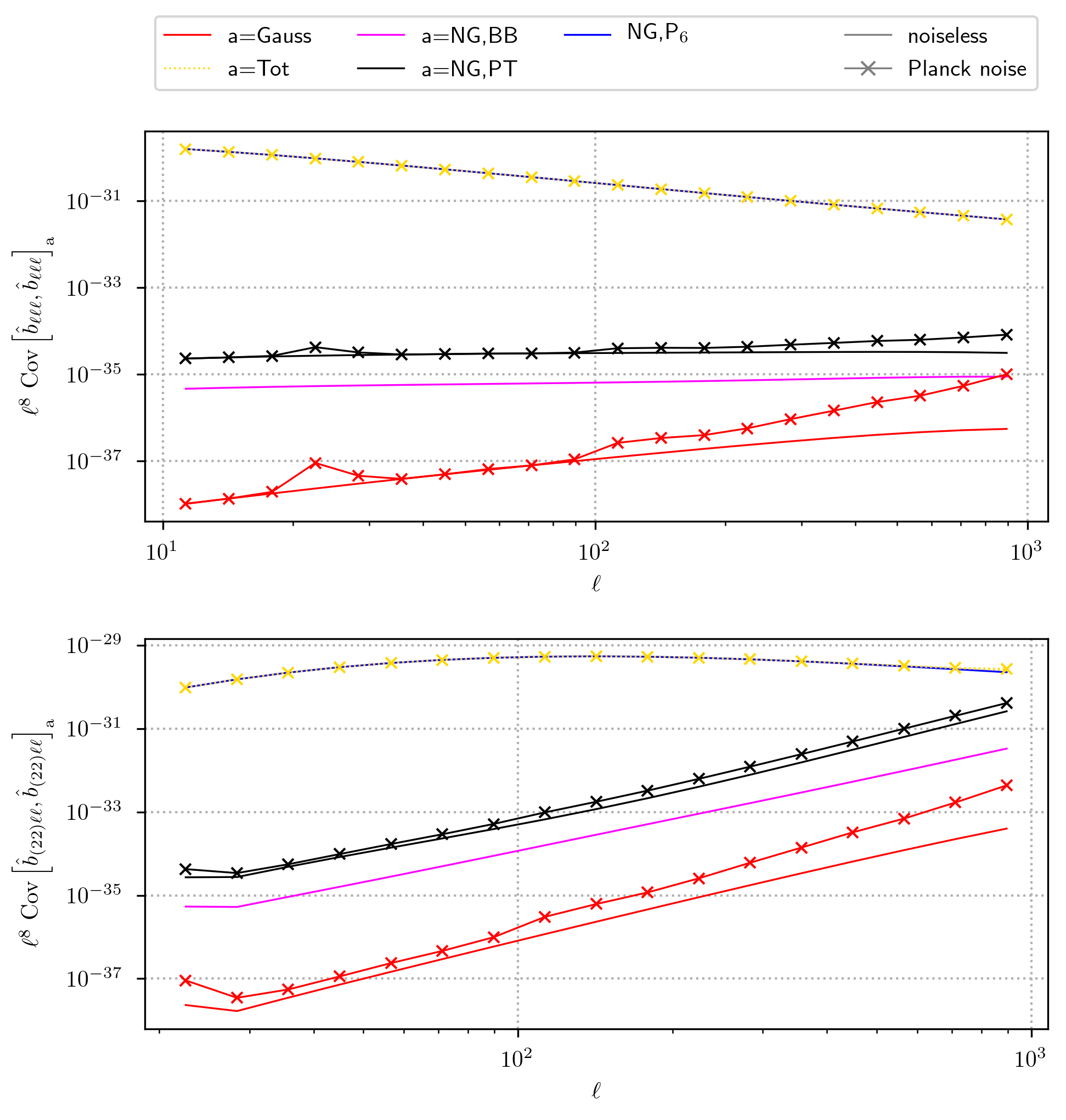}
    \caption{
    Selected components of the tSZ bispectrum covariance matrix.
    Whenever of relevance, the impact of the Gaussian Planck-like noise is displayed through makers. In the \textit{top} panel we focus on equilateral configuration while in the \textit{bottom} panel we focus on squeezed configuration of type $\{\sim22,\ell,\ell\}$. All the components are properly scaled for visualisation purpose.}
    \label{fig:CovarianceHierarchy_BS}
\end{figure}

In the right panel of figure \ref{fig:CovMat_BS} we show the fraction of the total covariance due to the 6-point connected term we just described, which is the dominant one:
as shown in figure~\ref{fig:CovarianceHierarchy_PS} and \ref{fig:CovarianceHierarchy_BS}, it is possible to identify a hierarchy among the different components listed above. The $\text{Cov}\left[\dots\right]_{\text{NG,T}}$ and the $\text{Cov}\left[\dots\right]_{\text{NG,6P}}$ component dominating the power spectrum and the bispectrum covariance, respectively, clearly emphasises the level of non-Gaussianities in the tSZ field.

\section{Power spectrum and bispectrum dependence on the parameters}
\label{sec:Derivatives}

It is well known that varying cosmological and gas parameter results in an amplitude shift of the power spectrum \cite[e.g.][]{Ade:2015ava, Bolliet:2017lha} and a slight tilt at smaller scales, where the power spectrum has a stronger dependence on the shape of the halo profile.
As a consequence, there are important degeneracies among the various parameters and, both with current and future data, one needs to consider additional observables to constrain various parameter jointly, either exploiting other datasets \cite[e.g.][]{Ade:2015ava} or other statistics extracted from the same tSZ maps \cite{Hurier:2017jgi}.

Overall, also the bispectrum displays a similar dependence on different parameters, leading again to an amplitude shift as the main effect.
This comes from the fact that tSZ correlators are dominated by the one-halo term, making the overall tSZ statistic nearly Poisson.
However, for the bispectrum we can also appreciate a non-negligible, configuration dependent modulation, which differs for varying parameters.

\begin{figure}
    \centering
    \includegraphics[width=\linewidth]{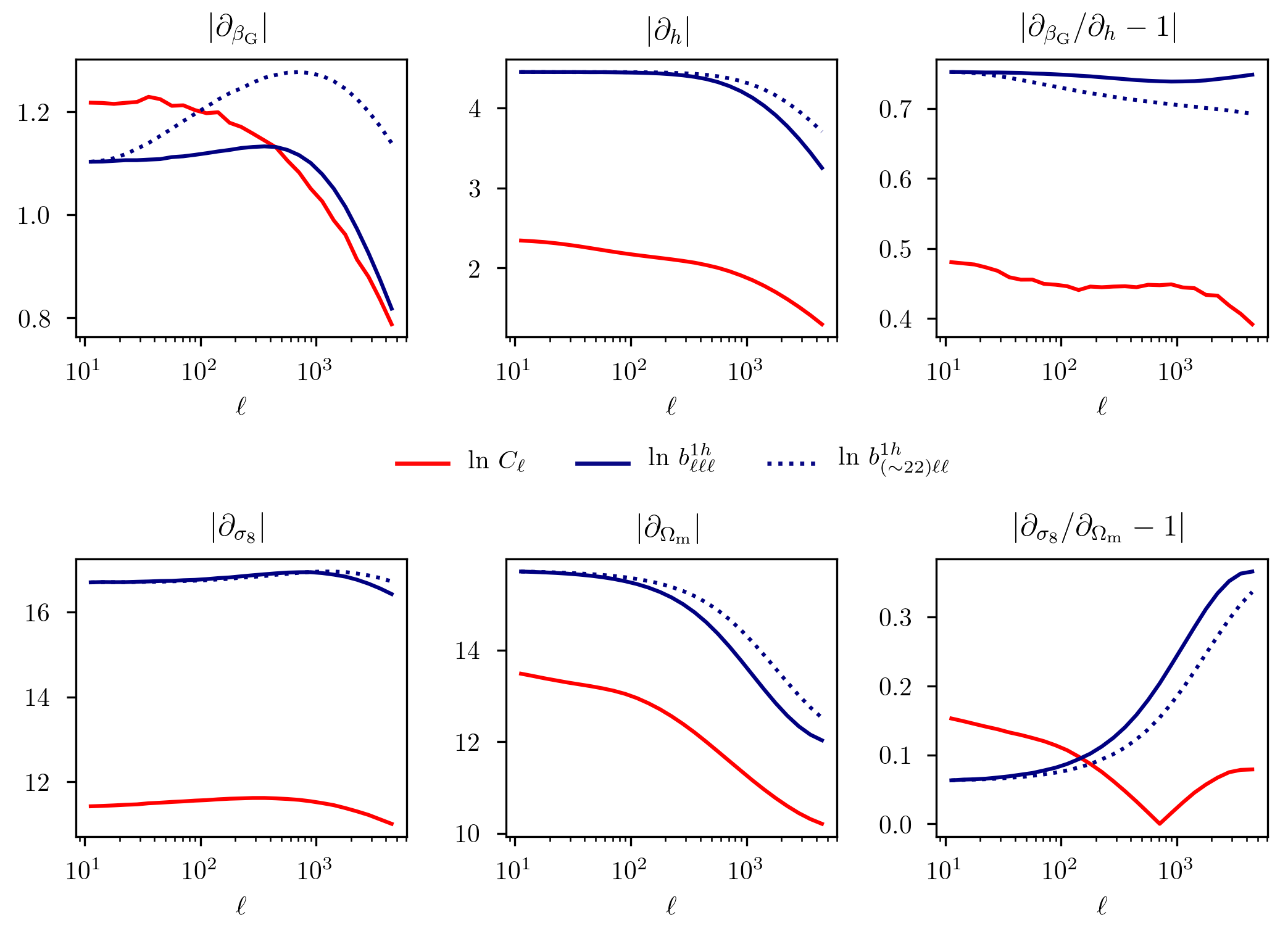}
    \caption{\textit{Left and central column}) logarithmic derivative of the power spectrum (red solid line) and of the bispectrum (dark blue line) with respect to different parameters of interest. Whenever the bispectrum is considered, we show equilateral configurations (solid line) and squeezed configurations of kind $\{\sim22,\ell,\ell\}$ (dotted line).  The parameters with respect to which the derivatives are taken are: $\beta_{\mathrm{G}}$ (\textit{top-left panel}), $h$ (\textit{top-central panel}), $\sigma_8$ (\textit{bottom-left panel}) and $\Omega_{\mathrm{m}}$ (\textit{bottom central panel}).  
    \textit{Right column}) fractional difference between the two derivatives on the left (per row). Color and line style code is the same as above.}
    \label{fig:EqSqBispectrumDerivatives}
\end{figure}

\begin{figure}
	\begin{subfigure}{0.7\textwidth}
		\includegraphics[width=\linewidth]{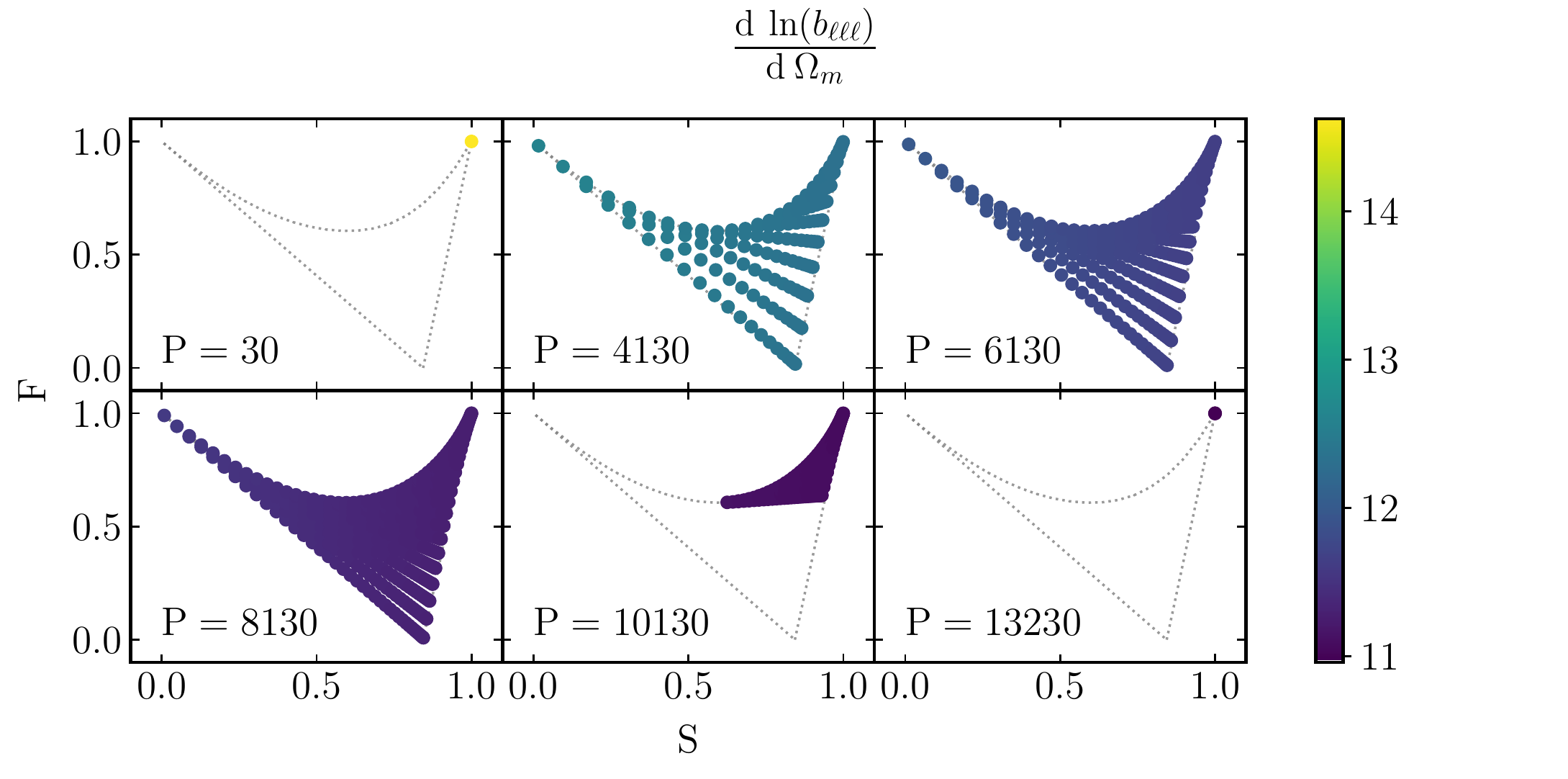} 
	\end{subfigure}
	\hspace{-1cm}
	\begin{subfigure}{0.35\textwidth}
		\vspace{0.3cm}
		\includegraphics[width=\linewidth]{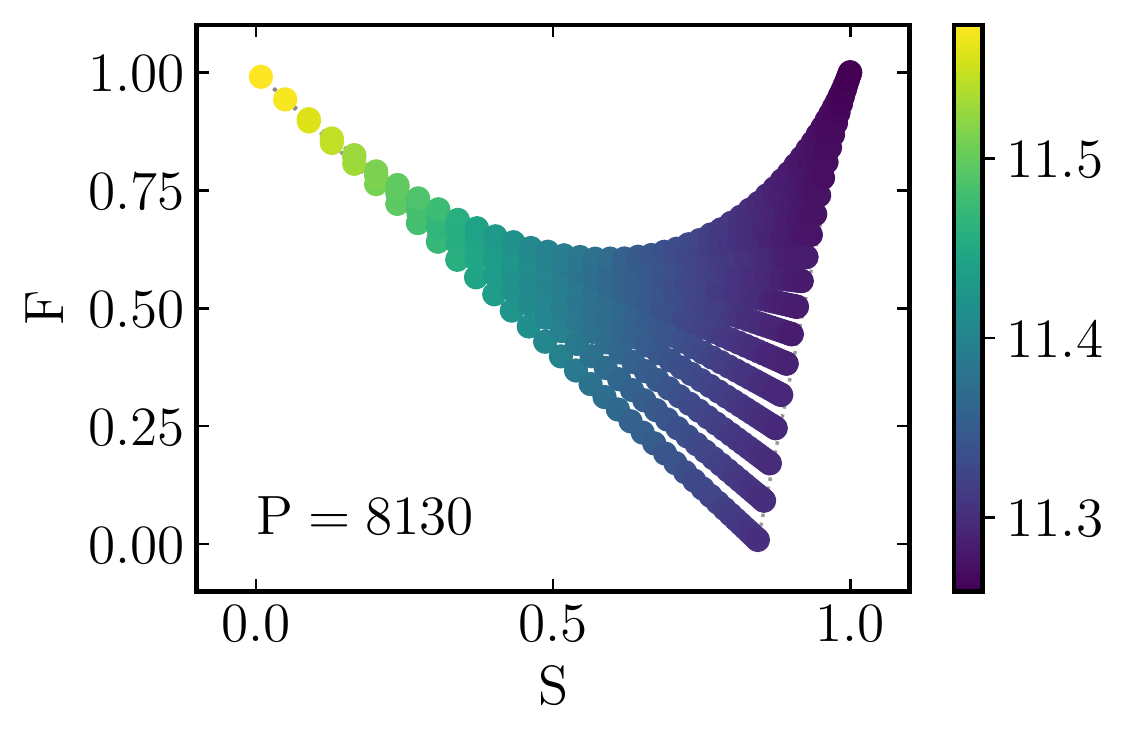}
	\end{subfigure}
	\begin{subfigure}{0.7\textwidth}
	\includegraphics[width=\linewidth]{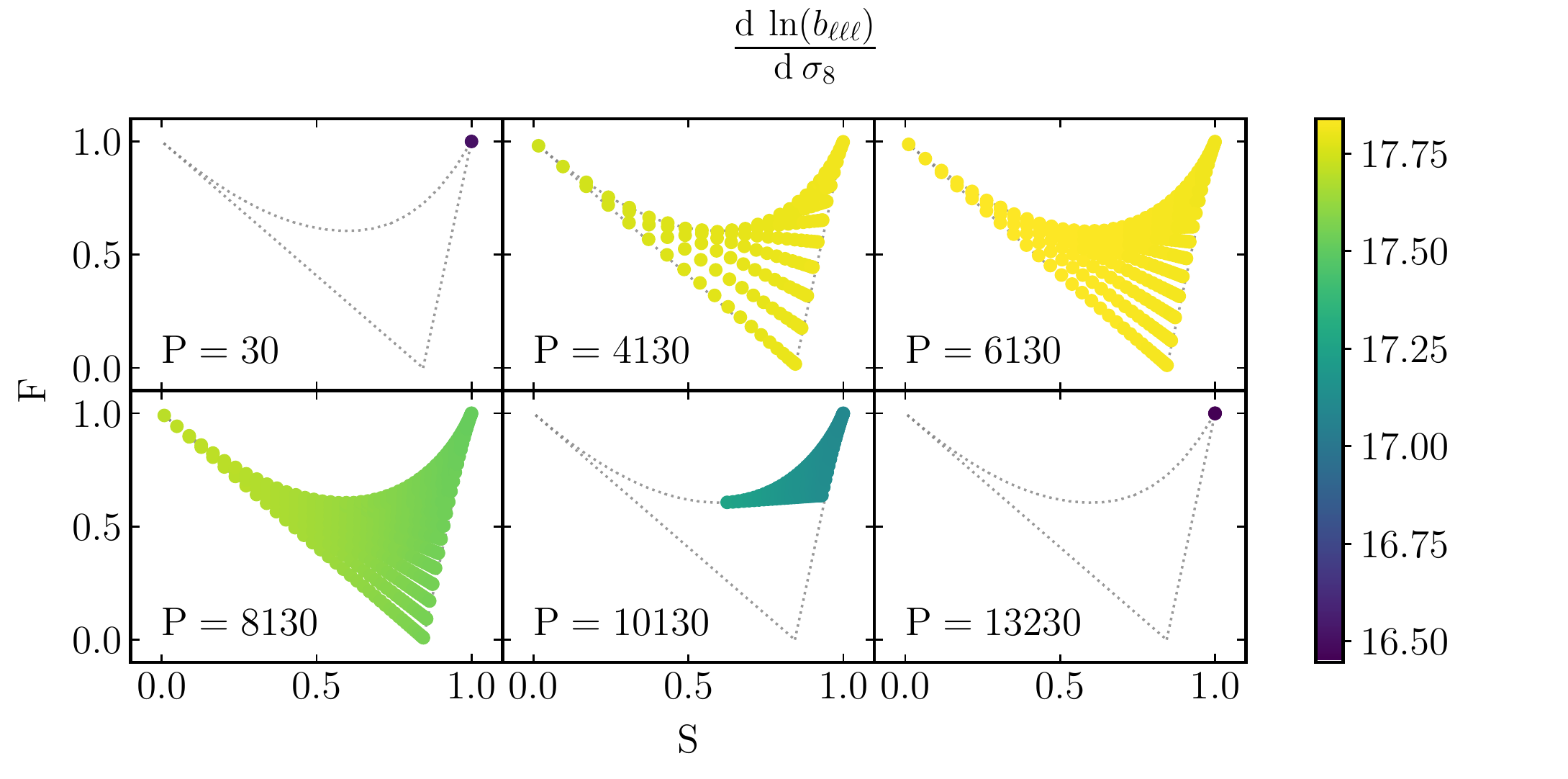} 
	\end{subfigure}
	\hspace{-1cm}
	\begin{subfigure}{0.35\textwidth}
		\vspace{0.3cm}
		\includegraphics[width=\linewidth]{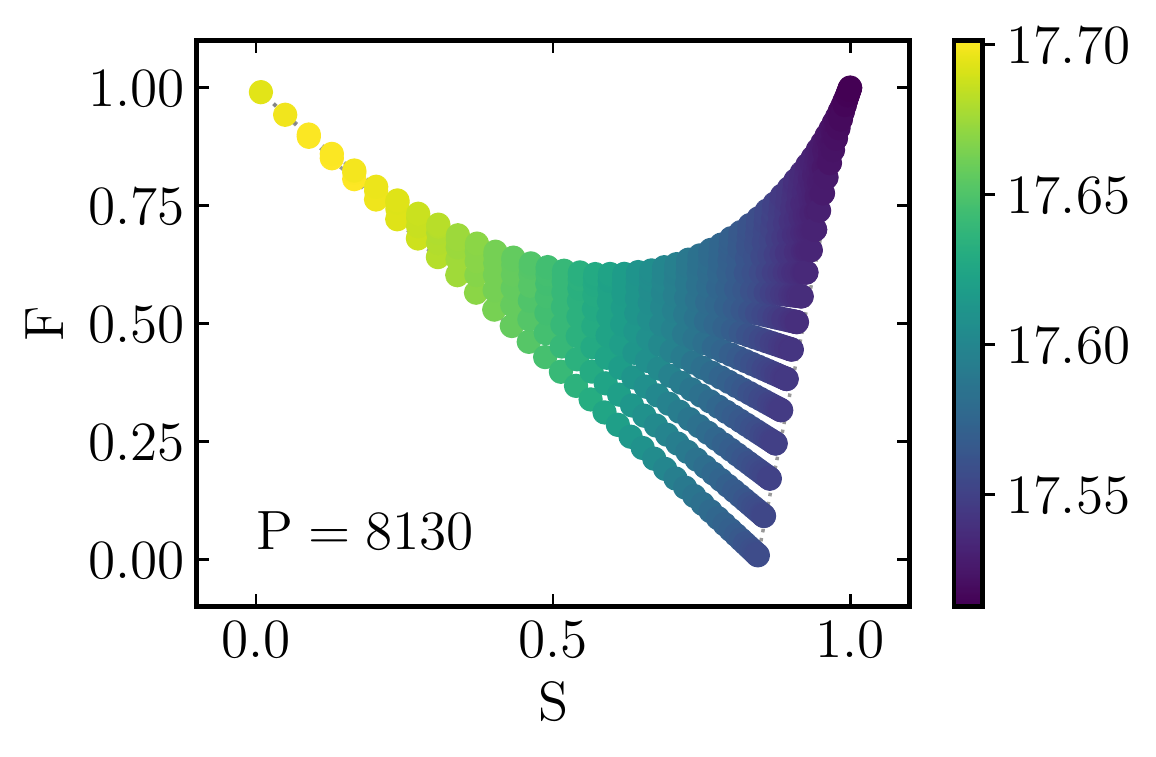}
	\end{subfigure}
	\begin{subfigure}{0.7\textwidth}
		\includegraphics[width=\linewidth]{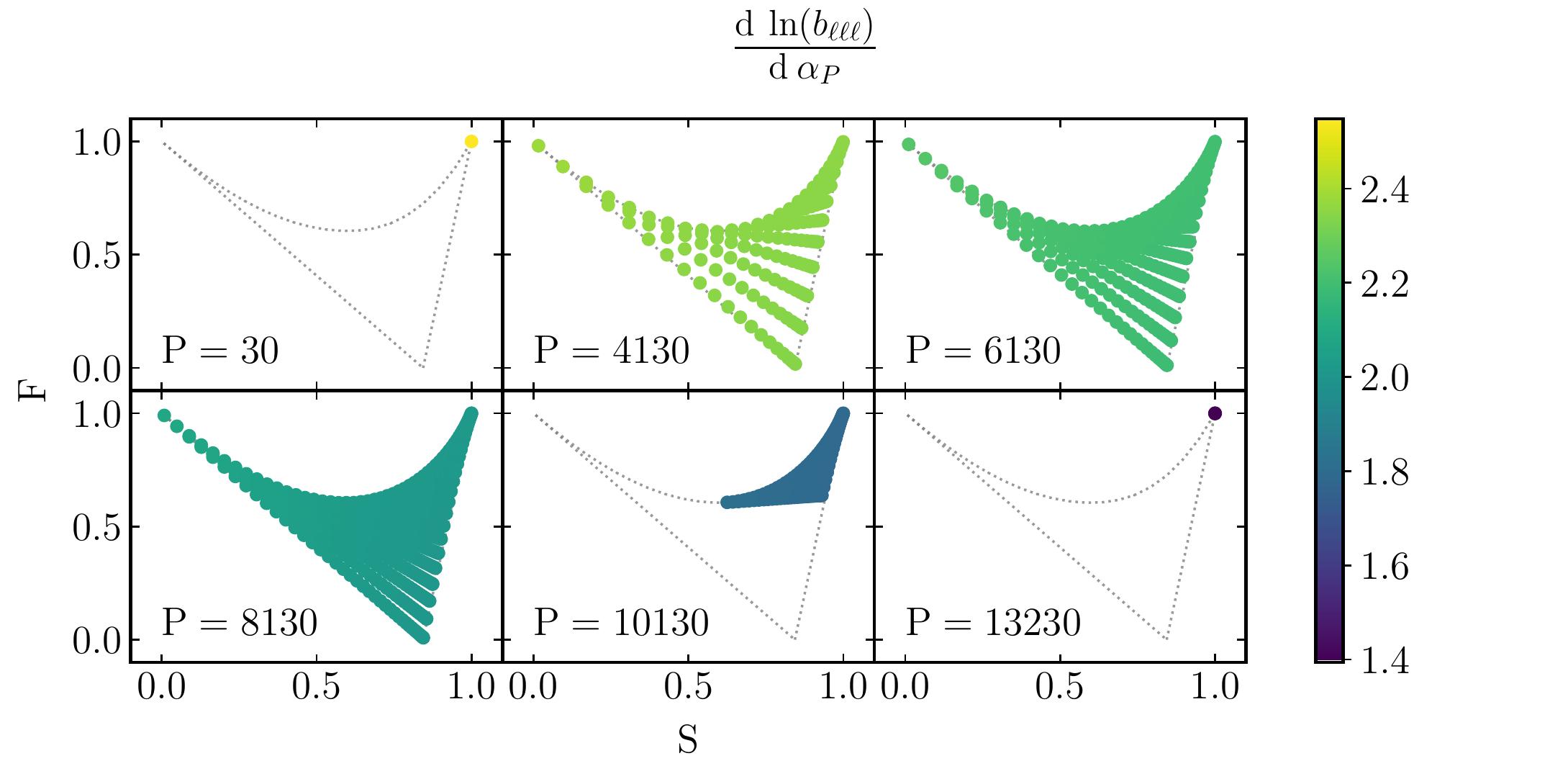} 
	\end{subfigure}
	\hspace{-1cm}
	\begin{subfigure}{0.35\textwidth}
		\vspace{0.3cm}
		\includegraphics[width=\linewidth]{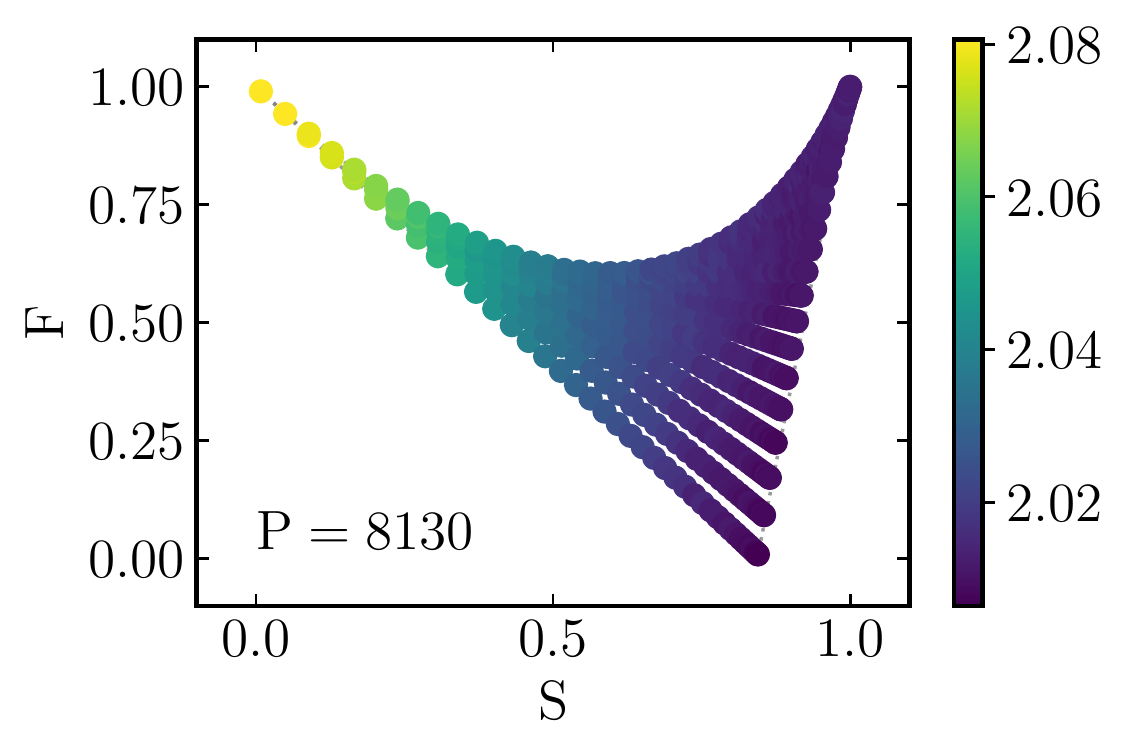}
	\end{subfigure}
    \caption{Logarithmic derivative of the bispectrum with respect to $\Omega_m$ (\emph{top}), $\sigma_8$ (\emph{middle}), and $\alpha_P$ (\emph{bottom}). On the right we also provide a rescaled plot for the perimeter value of P = $8130$ to illustrate the differences in derivatives for different triangles.}
    \label{fig:B_der_norm_Om_sig}
\end{figure}

Let us start from considering the bispectrum derivatives for the equilateral ($\ell_1=\ell_2=\ell_3$) and ``squeezed'' configurations ($\ell_1\approx 22,\ell_2=\ell_3$), and compare them to their power spectrum counterparts in figure \ref{fig:EqSqBispectrumDerivatives}.%
\footnote{We do not add the further requirement $\ell_{2,3}\gg \ell_1$, so not all the configurations in the squeezed set are technically ``squeezed''.}
Much like the power spectrum, the derivatives of the equilateral bispectrum are similar to each other, most of them being somewhat flat up to $\ell\approx 10^3$ where the halo inner structure starts to be resolved, with a subsequent reduction (in absolute value) on smaller scales.
On the other hand, the derivatives of the squeezed bispectrum have a slightly more variegated phenomenology among themselves and, on top of that, they differ from the equilateral bispectrum ones.
For this reason we can intuitively expect squeezed configurations to be less affected by parameter degeneracies.
While in section \ref{sec:EqSqAnalyis} we will confirm this expectation, we will also find that the conditional errors will be much bigger to begin with than what we find using only equilateral configurations, further motivating the analysis of the full bispectrum.
Using the bispectrum parametrization introduced in \cite{Lacasa:2011ej},
\begin{equation}
\begin{split}
    P &\equiv \ell_1 + \ell_2 + \ell_3 \, , \qquad
    F \equiv \frac{32(\tilde{\sigma}_2-\tilde{\sigma}_3)}{3} + 1 \, , \qquad
    S \equiv \tilde{\sigma}_3 \, ,
\\
    \tilde{\sigma}_2 &\equiv 12 \frac{(\ell_1\ell_2 + \ell_1\ell_3 + \ell_2\ell_3)}{(\ell_1+ \ell_2 + \ell_3)^2}-3 \, ,
    \qquad
    \tilde{\sigma}_3 \equiv 27\frac{\ell_1\ell_2\ell_3}{(\ell_1+ \ell_2 + \ell_3)^3}\, ,
\end{split}
\end{equation}
in figure
\ref{fig:B_der_norm_Om_sig}
we show how each configuration responds to parameter changes in a slightly different manner.
Even though each one has lower significance than the power spectrum, this means that it is easier to discern the effects of the various parameter changes.
We will quantify this statement in section \ref{sec:ForecastResults}.
The reason why different bispectrum configurations respond differently to changes in the various parameters lies in the slightly different ranges of halo masses and redshifts that contribute the most to the specific configuration.
This is investigated in appendix~\ref{sec:A_BispectrumKernel}.

\section{Gaussian experimental noise and foregrounds}
\label{sec:GaussianForegrounds}

To assess the impact of instrumental noise and foregrounds on the forecast we employ the methodology of \cite{Cooray:2000xh, Hill:2013baa}.
We assume to use multi-channel measurements to run a simple internal linear combination (ILC) to separate foregrounds from the signal.
To do so, we will assume perfect knowledge of the foreground spectral energy densities, which might be unrealistic in a real life scenario \cite{Chluba:2017rtj,Abitbol:2017vwa,Rotti:2020tlo}.
We leave the assessment of the impact of more refined foreground modelling to future work.

\subsection{Spectral components}
A comprehensive list of the spectral components that affect tSZ measurements 
is given by
instrumental noise, CMB, free-free, thermal dust, synchrotron, radio point sources, infrared point sources.
We will consider all of these contributions even if we use approximate descriptions that are anyway commonly used in the literature.

We assume we can factorize the inter-frequency correlation in three ingredients: the spectral energy density expressed with respect to the CMB black-body $\Theta^\text{sc}(\nu)$, a spectral coherence factor that parametrize the correlation between different instrumental frequency channels $R(\nu_i, \nu_j, \xi^\text{sc})$, and an angular scaling $C_\ell^\text{sc}$:
\begin{equation}
    C^\text{sc}(\nu_i,\nu_j) 
    = 
    \Theta^\text{sc}(\nu_i)\Theta^\text{sc}(\nu_j)
    R(\nu_i, \nu_j, \xi^\text{sc}) C_\ell^\text{sc} \, .
\end{equation}
In \cite{Tegmark:1999ke} the spectral coherence in two channels $\nu_i$ and $\nu_j$, for each component, was parametrized as 
\begin{equation}
    R(\nu_i, \nu_j, \xi^\text{sc}) = \exp
    \left[
        -\frac{1}{2}
        \left(
            \frac{\log(\nu_i)-\log(\nu_j)}{\xi^\text{sc}}
        \right)^{\!2} \,
    \right],
\qquad
    \xi^\text{sc}\approx\frac{1}{\sqrt{2}\Delta \alpha^\text{sc}} \, ,
\end{equation}
where, for each spectral component, $\Delta \alpha^\text{sc}$ is the variance of the spectral index over the sky.
Depending on the spectral component, $\xi^\text{sc}$ varies in the range $[0,\infty)$.
In particular, we assume the instrument frequency channels to be independent ($\xi^\text{sc} \rightarrow 0$), and notice that the CMB has perfect correlation across all the frequencies ($\xi^\text{sc} \rightarrow \infty$).

$\Theta^\text{sc}$, $\Delta\alpha^\text{sc}$, and $C_\ell^\text{sc}$ vary for each spectral component.
We will use the values and functional forms provided in  \cite{Abitbol:2017vwa, Hill:2013baa, Tegmark:1999ke, Dunkley:2013vu}, that we report in appendix \ref{app:Foregrounds_spectra_shape}.

\subsection{Component separation}
The ILC aims at recovering, multipole by multipole, a signal of known spectral dependence computing a weighted average of data collected at different frequencies.%
\footnote{All quantities in this section (except $N_\ell$) are calculated separately for each $(\ell,m)$.}
The weights are calculated in such a way to have unitary response to the desired spectral shape, while minimizing spurious contributions from other spectral components.

The internal linear combination weights and the effective noise term after multifrequency subtraction are respectively \cite{Cooray:2000xh}
\begin{equation}
    \vec{w}=\frac{\left[\sum_\text{sc} C^\text{sc}(\nu_i,\nu_j)\right]^{-1} \, \vec{g}}{\vec{g} \cdot \left[\sum_\text{sc} C^\text{sc}(\nu_i,\nu_j)\right]^{-1} \, \vec{g}} \, ,
\quad
    N= \vec{g} \cdot 
    \left[
        \sum \nolimits_\text{sc} C^\text{sc}(\nu_i,\nu_j)
    \right]^{-1} \vec{g} \, .
\label{eq:ILC_weights_noise}
\end{equation}
where $\vec{g}=(g(\nu_1),\cdots,g(\nu_1))^T$ is a vector of the value of the tSZ spectral shape evaluated on the observed channels.
This result, valid for a single multipolar coefficient, for each $\ell$ can be averaged over all $m$: we can therefore express the effective noise term $N_\ell = N/(2\ell+1)$.%
\footnote{As a sanity check we successfully reproduced figure 12 of \cite{Hill:2013baa}.
To do so one has to use $N_\ell^\text{sc}= \vec{w} \cdot \mathbb{C}^\text{sc} \vec{w}$, where the $\vec{w}$ are still calculated according to eq. \eqref{eq:ILC_weights_noise}.}

\section{Forecast results}
\label{sec:ForecastResults}
 We now have all the ingredients to understand what is the information content of the power spectrum, of the full bispectrum, and of their combination.
We use a Fisher forecast to estimate how well a particular experiment can constrain a parameter through some observed quantity. The components of the Fisher information matrix are defined as \cite{Heavens:2009nx}:
\begin{equation}\label{eq:general_Fisher}
	F_{ij} = 
	\left\langle 
		- \frac{\partial^2( ln\,\mathcal{L} )}{\partial \theta_i\,\partial \theta_j} 
	\right\rangle,
\end{equation}
where $\mathcal{L}$ is the likelihood function and the $\theta_i$ are the parameters we want to constrain. 

In case of Gaussian distributed data and rotational invariance of the observable, the Fisher matrix can also be written as \cite{Carron:2012pw}: 
\begin{equation}\label{eq:Fisher}
	F_{ij} = 
	\frac{\partial \mathcal{O}}{\partial {\theta_i}}^T
	{\rm Cov}^{-1} \,
	\frac{\partial \mathcal{O}}{\partial {\theta_j}}\,
	\: + \:
	\frac{1}{2} \: {\rm Tr} \left[
	\,{\rm Cov}^{-1} \,
	\frac{\partial\,{\rm Cov}}{\partial\theta_i} 
	\,{\rm Cov}^{-1} \,
	\frac{\partial\,{\rm Cov}}{\partial\theta_j}
	\right]
\end{equation}
where ${\rm Cov}={\rm Cov}[\mathcal{O}, \mathcal{O}]$ is the covariance matrix, described in section \ref{3.2}.
We drop the second term as all the relevant information is contained in the mean of the observable, while the covariance dependence on the parameters would introduce spurious information that cannot be extracted by the considered estimator \cite{Carron:2012pw}.
In principle neither the power spectrum nor the bispectrum follow a Gaussian distribution, but we expect it to be a good approximation for the binned spectra in virtue of the central limit theorem.
This ansatz should however be validated against simulations or more refined analysis \citep{DALI,DALII} that we leave for future work.

For each parameter the conditional error is:
\begin{equation}
	\sigma_{i,\,{\rm conditional}} = \frac{1}{\sqrt{F_{ii}}} \, ,
\end{equation}
which is the error one gets fitting only one parameter while fixing all the others.
What we are more interested in, however, is the marginalized error.
The diagonal entries of the inverse of a Fisher matrix give the error for each parameter marginalized over all the others, which one would obtain by a multivariate fit:
\begin{equation}
	\sigma_{i} = \sqrt{{\rm Cov}(\theta_i,\theta_i)} = \sqrt{(F^{-1})_{ii}}\,.
\end{equation}
The off diagonal elements then give us the covariance between two parameters, ${\rm Cov}(\theta_i,\theta_j)$.

We start by considering an ideal, noiseless experiment in the absence of foregrounds, to understand the general properties of the spectra in a simplified case, then we show a forecast for the already available \textit{Planck} data, and finally we consider two realistic future surveys.

\subsection{Noiseless survey with no foreground contamination}
\label{sec:Results:CVL}

In order to set an upper limit on the amount of information that can be in principle extracted from the tSZ power spectrum and bispectrum, we start by considering the case of a noiseless survey while also neglecting any issue of foreground contamination. This will also help us to build understanding of the main factors which affect the final constraints.
For brevity we will refer to this configuration, slightly improperly, as Cosmic Variance Limited (CVL) survey.
This initial oversimplified analysis will then be generalized in the following sections, including all realistic effects.

In tables \ref{tab:results_noiseless_ell1000} and \ref{tab:results_noiseless_ell5000} we show the forecasted error bars on all the parameters for a full sky survey, with $\ell_\text{max}= 1000$ and $\ell_\text{max}= 5000$ respectively.
To assess the impact of a change in $\ell_\text{min}$ instead, we repeated both analysis switching from $\ell_\text{min} = 10$ to $\ell_\text{min}= 70$ and found quantitatively negligible differences.

\begin{table}
    \centering
    \begin{tabular}{|c|ccc|ccc|}
    \hline
    & \multicolumn{3}{c|}{Conditional} & \multicolumn{3}{c|}{Marginalized}\\
                & PS $1\sigma$ & BS $1\sigma$ & PS$\oplus$BS $1\sigma$ & PS $1\sigma$ & BS $1\sigma$ & PS$\oplus$BS $1\sigma$\\
    \hline
$\Omega_m$	&	0.0021	&	0.0092	&	0.0015	&	8.1	&	0.12	&	0.092	\\
$h$	&	0.013	&	0.033	&	0.010	&	16	&	0.36	&	0.34	\\
$\sigma_8$	&	0.0020	&	0.0075	&	0.0015	&	3.4	&	0.047	&	0.047	\\
$n_S$	&	0.060	&	0.060	&	0.051	&	8.0	&	0.43	&	0.30	\\
$w_0$	&	0.096	&	0.31	&	0.071	&	105	&	0.59	&	0.54	\\
$b_\text{HSM}$	&	0.0060	&	0.023	&	0.0043	&	407	&	0.93	&	0.81	\\
$P_0$	&	0.097	&	0.36	&	0.071	&	8212	&	14	&	12	\\
$\alpha_P$	&	0.12	&	0.071	&	0.069	&	60	&	0.20	&	0.18	\\
$c_{500,c}$	&	0.0059	&	0.024	&	0.0041	&	4.4	&	0.065	&	0.064	\\
$\alpha_G$	&	0.0041	&	0.015	&	0.0029	&	64	&	0.065	&	0.064	\\
$\beta_G$	&	0.023	&	0.030	&	0.014	&	1.8	&	0.034	&	0.034	\\
$\gamma_G$	&	0.011	&	0.034	&	0.0085	&	245	&	0.24	&	0.20	\\
    \hline
    \end{tabular}
    \caption{Conditional and marginalized error forecasts for a CVL experiment with  $\ell_\text{max}=1000$ and $f_\text{sky}=1$.}
    \label{tab:results_noiseless_ell1000}
\end{table}

\begin{table}
    \centering
    \begin{tabular}{|c|ccc|ccc|}
    \hline
    & \multicolumn{3}{c|}{Conditional} & \multicolumn{3}{c|}{Marginalized}\\
                & PS $1\sigma$ & BS $1\sigma$ & PS$\oplus$BS $1\sigma$ & PS $1\sigma$ & BS $1\sigma$ & PS$\oplus$BS $1\sigma$\\
    \hline
$\Omega_m$	&	0.00044	&	0.0017	&	0.00031	&	0.86	&	0.025	&	0.018	\\
$h$	&	0.0040	&	0.0079	&	0.0033	&	5.7	&	0.083	&	0.076	\\
$\sigma_8$	&	0.00040	&	0.0013	&	0.00030	&	0.43	&	0.011	&	0.010	\\
$n_S$	&	0.0050	&	0.026	&	0.0030	&	0.77	&	0.080	&	0.063	\\
$w_0$	&	0.0052	&	0.014	&	0.0039	&	11	&	0.10	&	0.092	\\
$b_\text{HSM}$	&	0.0013	&	0.0044	&	0.00091	&	78	&	0.16	&	0.16	\\
$P_0$	&	0.018	&	0.060	&	0.014	&	1754	&	2.7	&	2.6	\\
$\alpha_P$	&	0.0050	&	0.022	&	0.0030	&	2.81	&	0.040	&	0.037	\\
$c_{500,c}$	&	0.0014	&	0.0054	&	0.00097	&	0.90	&	0.016	&	0.015	\\
$\alpha_G$	&	0.00082	&	0.0026	&	0.00059	&	7.84	&	0.015	&	0.014	\\
$\beta_G$	&	0.0063	&	0.0067	&	0.0036	&	0.31	&	0.0077	&	0.0077	\\
$\gamma_G$	&	0.0017	&	0.0045	&	0.0013	&	50	&	0.039	&	0.036	\\
    \hline
    \end{tabular}
    \caption{Conditional and marginalized error forecasts for a CVL experiment with  $\ell_\text{max}=5000$ and $f_\text{sky}=1$.}
    \label{tab:results_noiseless_ell5000}
\end{table}


In both cases, if one considers the conditional errors, the power spectrum outperforms the bispectrum analyzed on its own by a factor $2\sim 10$;
obviously when the two signals are analyzed jointly, the errors shrink marginally with respect to the power spectrum case.
If all but one parameters are perfectly known, the power spectrum, comparatively bigger than its noise with respect to the bispectrum, leads to tighter constraints on the last parameter.
This outcome is overturned by the marginalization needed for a joint fit of the parameters.

\begin{figure}
    \centering
    \includegraphics[width=\linewidth]{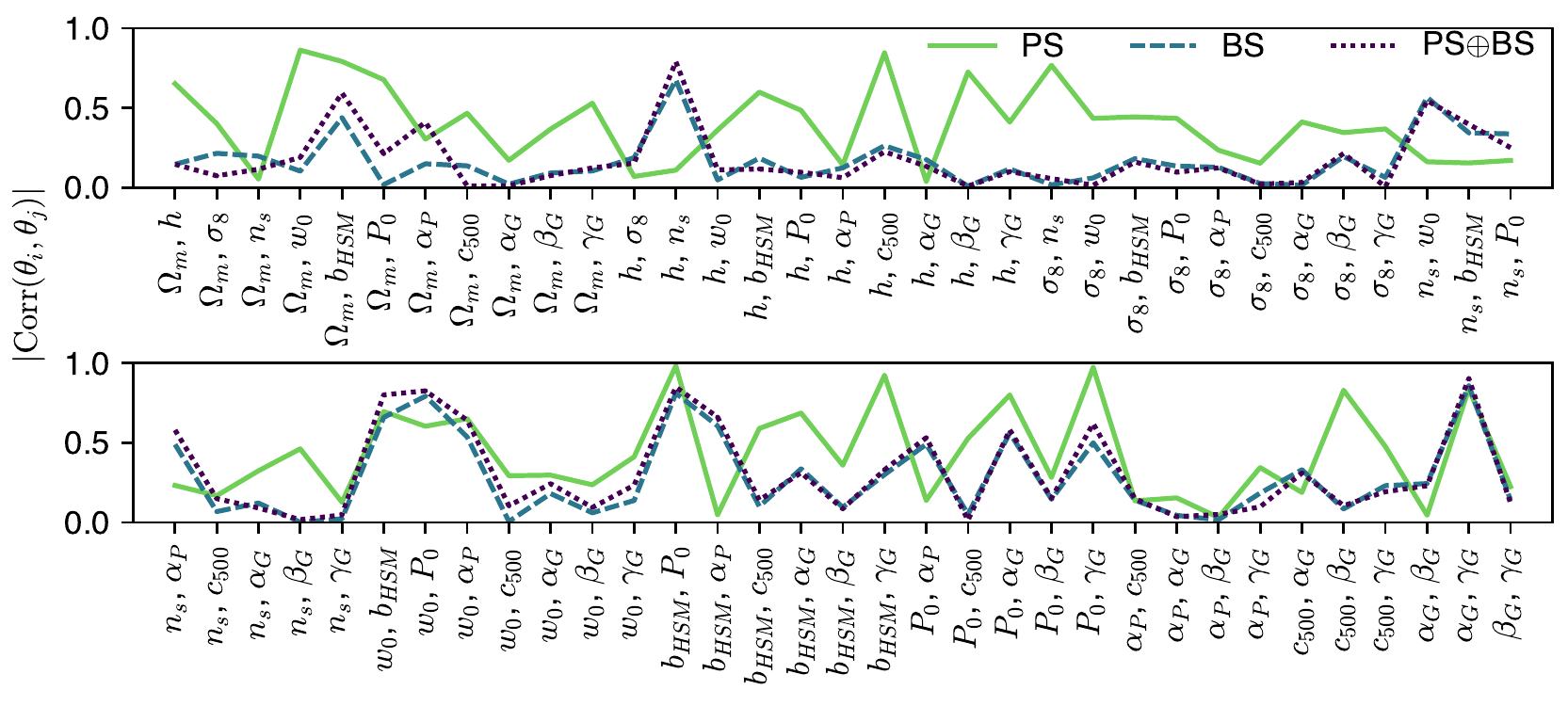}
    \caption{Comparison of the Pearson correlation coefficients for all couples of model parameters.
    Broadly speaking the correlations among the parameters are clearly higher when one considers the power spectrum with respect to the bispectrum.}
    \label{fig:Pearson_NoNoiseNoForeground5000}
\end{figure}

\begin{figure}
    \centering
    \includegraphics[width=.5\linewidth]{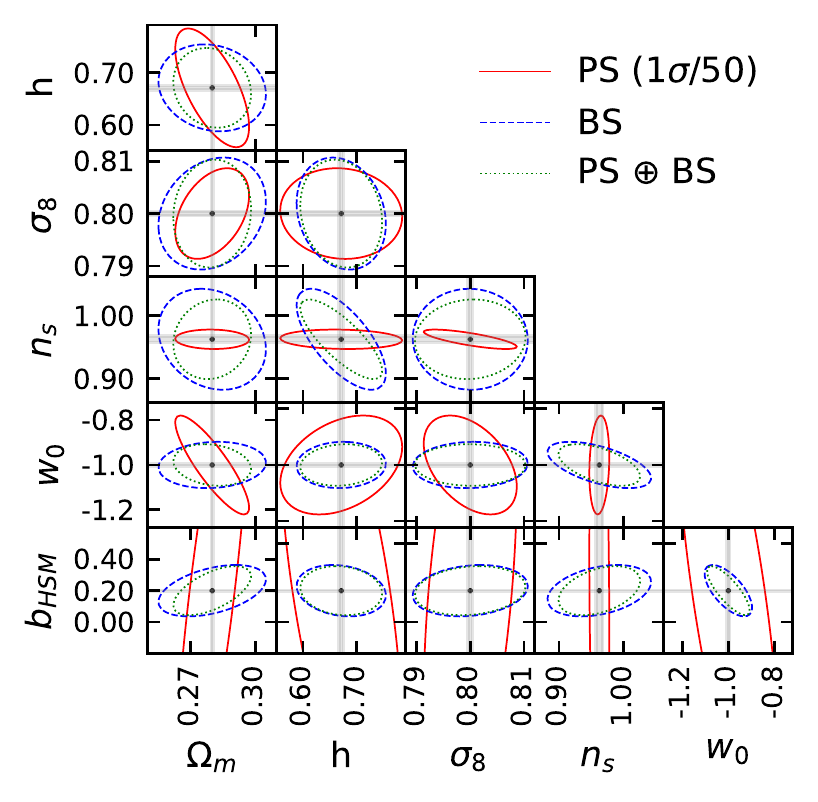}%
    \includegraphics[width=.5\linewidth]{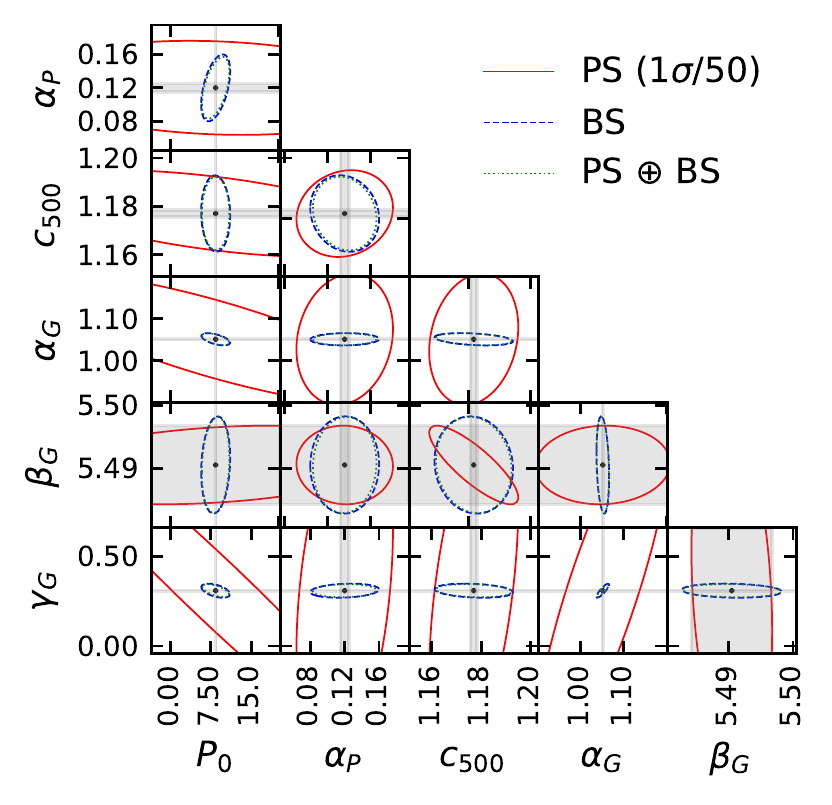}
    \caption{Triangle plots for cosmological (\textit{left}) and gas parameters (\textit{right}). Here we consider a Cosmic variance limited experiment with perfect foreground separation and $\ell_\text{max}=5000$. Notice that the power spectrum $1\sigma$ ellipses have been rescaled by a factor to fit in the same graph.
    The grey bands are the power spectrum conditional errors.}
    \label{fig:NoNoiseNoForeground5000}
\end{figure}

As anticipated in section \ref{sec:Derivatives}, at power spectrum level the parameters have serious degeneracies among all of them, whereas the bispectrum has a more diverse response to parameters change.
To quantify this statement, in figure \ref{fig:Pearson_NoNoiseNoForeground5000} we compare for each couple of parameters the Pearson correlation coefficient
\begin{equation}
    \text{Corr}(\theta_i, \theta_j)
    \equiv
    \frac{\text{Cov}(\theta_i, \theta_j)}
    {\sqrt{\text{Cov}(\theta_i, \theta_i) \, \text{Cov}(\theta_j, \theta_j)}} \, , 
\end{equation}
that one obtains using the power spectrum, the bispectrum, and their combination, in the case of $\ell_\text{max}= 5000$.
Correlations are generally much higher for the power spectrum and for this reason after marginalization the power spectrum loses much of its constraining power.
The same information is also displayed in figure \ref{fig:NoNoiseNoForeground5000}, where we show the triangle plot of cosmological and gas parameters.

\begin{figure}
    \centering
    \includegraphics[width=.5\linewidth]{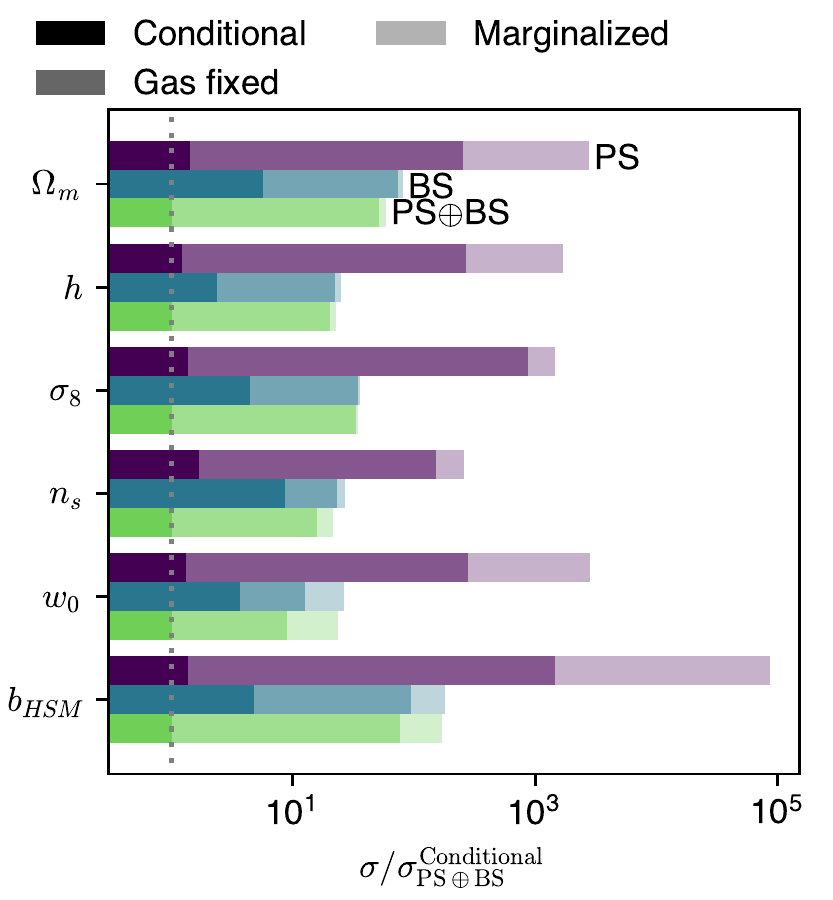}%
    \includegraphics[width=.5\linewidth]{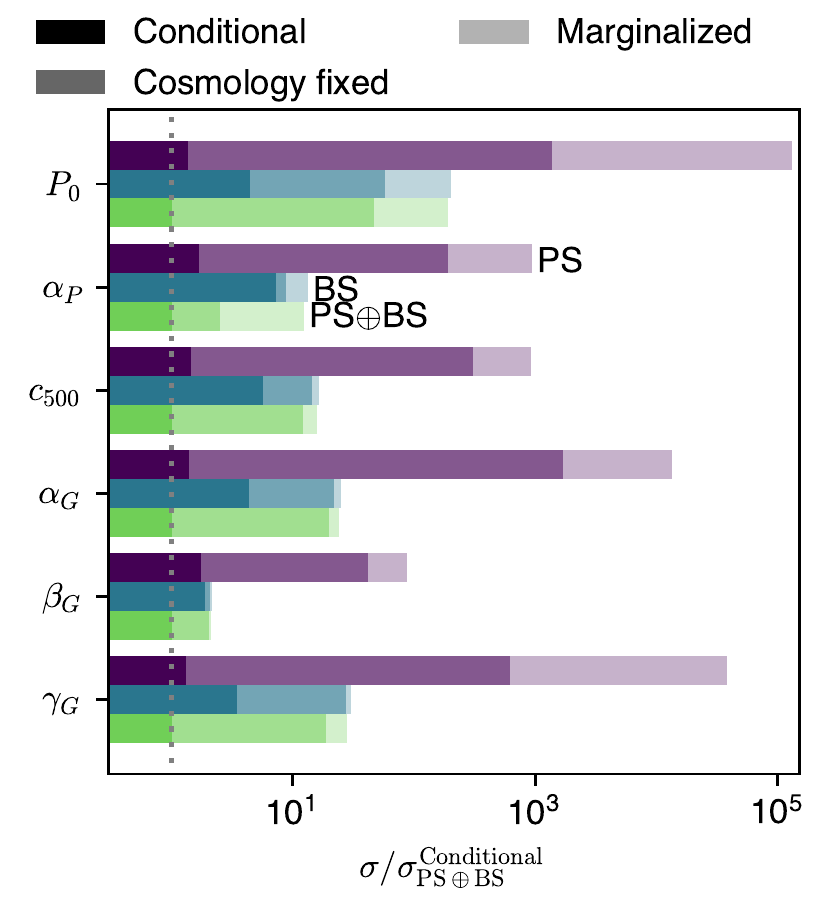}
    \caption{Impact of degeneracies on the parameters error bars.
    \emph{Left}) for each cosmological parameter we show the error marginalized over all others parameter, the error marginalized over cosmological parameters, fixing the gas ones, and the conditional error; all divided by the conditional error of power spectrum and bispectrum combined.
    This is repeated for (top to bottom in each triplet of bars) power spectrum, bispectrum, and combination of the two.
    The constraining power of the power spectrum is hindered by the marginalization over other parameters; on the other hand the bispectrum, in principle less constraining, is less affected.
    \emph{Right}) same, but switching cosmological and gas parameters.
    }
    \label{fig:ErrorBars_NoNoiseNoForeground5000}
\end{figure}

To show the dramatic impact of degeneracies and the incremental reduction of constraining power when more and more parameters are jointly fitted, in figure \ref{fig:ErrorBars_NoNoiseNoForeground5000} we directly compare the error bars recovered in various scenarios.
For each parameter the baseline (tightest possible constraints in our analysis) is the conditional error with a joint power spectrum and bispectrum fit.
All other errors are shown in figure \ref{fig:ErrorBars_NoNoiseNoForeground5000} as ratio to this value.
For each parameter the top error bar is calculated with the power spectrum analysis, the middle one with bispectrum analysis and the bottom one with the joint fit of the two.
For each bar the most saturated and least saturated color show the conditional and marginalized error, respectively.
In the left panel the mid-tone bar is obtained fixing the value of all gas parameters and marginalizing over cosmological ones;
in the right panel we did the opposite.

Having made the point that the bispectrum allows to separately fit multiple parameters, we now inquire what are the prospects considering tSZ observations combined to prior knowledge of cosmological and gas parameters.
Generally, the gas parameters are fixed to the best-fit values of simulations \cite{Battaglia:2011cq}, external X-Ray \cite{Arnaud:2009tt}, or stacked tSZ clusters \cite{Ade:2012ngj} measurements, and not varied throughout the analysis of cosmological parameters.
On the other hand, one can also think of using the \textit{Planck} primary anisotropies measurements to set a prior on the cosmological parameters, and exploit tSZ anisotropies data to cross-check the gas parameters.
We explore both strategies in table \ref{tab:results_noiseless_ell5000_Prior}.
In the left side we fixed $(P_0, \alpha_P, c_{500,c}, \alpha_G, \beta_G, \gamma_G)$, removing the respective rows and columns from the Fisher matrix.
In the right side, instead, we add a (Gaussian) prior to the cosmological parameters, obtained from the $1\sigma$ errors in the column TT,TE,EE+lowE+lensing+BAO in table 2 of \cite{Aghanim:2018eyx}, with the additional assumption of the errors being uncorrelated.
Since our paper is meant to be a proof of concept, and a comparison between the power spectrum and the bispectrum, we refrain from comparing the results from table \ref{tab:results_noiseless_ell5000_Prior} with state of the art observations for both cosmological and gas parameters, as our forecast are bound to be an upper limit and most likely over-optimistic.
However, they serve as a comparison between the power spectrum and the bispectrum efficiency and to motivate the analysis of both.

\begin{table}
    \centering
    \begin{tabular}{|c|ccc|c|ccc|}
    \hline
    & \multicolumn{3}{c|}{Gas fixed} & & \multicolumn{3}{c|}{\textit{Planck} prior}\\
                & PS $1\sigma$ & BS $1\sigma$ & PS$\oplus$BS $1\sigma$ & & PS $1\sigma$ & BS $1\sigma$ & PS$\oplus$BS $1\sigma$\\
    \hline
$\Omega_m$ &	0.078	&	0.023	&	0.016	&	$P_0$	&	1230	&	2.2	&	2.0	\\
$h$ &	0.89	&	0.074	&	0.068	&	$\alpha_P$	&	1.5	&	0.029	&	0.012	\\
$\sigma_8$ &	0.26	&	0.010	&	0.010	&	$c_{500,c}$	&	0.39	&	0.015	&	0.015	\\
$n_S$ &	0.45	&	0.069	&	0.047	&	$\alpha_G$	&	6.9	&	0.014	&	0.014	\\
$w_0$ &	1.1	&	0.050	&	0.035	&	$\beta_G$	&	0.15	&	0.0076	&	0.0075	\\
$b_\text{HSM}$ &	1.3	&	0.086	&	0.069	&	$\gamma_G$	&	41	&	0.039	&	0.036	\\
    \hline
    \end{tabular}
    \caption{
    \emph{Left}) marginalized error forecasts for cosmological parameters assuming the gas parameters are fixed.
    \emph{Right}) marginalized error forecast for gas parameters using \Planck priors on the cosmological parameters.
    In this table we always considered a CVL experiment with $\ell_\text{max}=5000$ and  $f_\text{sky}=1$.}
    \label{tab:results_noiseless_ell5000_Prior}
\end{table}

Our results heavily depend, through the covariance, on high order correlations functions.
In particular we study the impact of massive halos at low redshift, since higher order correlators are incrementally sensitive to the signal from these clusters.
In principle, to calculate the $y$-distortion along the line of sight one has to integrate from 0 to infinity,
and again in principle the HMF has been calibrated in \cite{2010ApJ...724..878T} integrating clusters up to a mass of $10^{16} M_\odot$.
However, the most massive cluster in the \textit{Planck} SZ cluster catalogue \cite{Ade:2015gva} has a mass of $1.61 \times 10^{15} M_\odot$, whereas the closest detected cluster is at $z=0.011$.
To account for these observational constraints we repeat our analysis enforcing these values as upper limits in the redshift and mass integrations.
The results that follow are shown in table \ref{tab:results_noiseless_ell5000_zMcut} which is the main result of the paper.

Removing massive nearby clusters has a deeper impact on higher order correlators with respect to the power spectrums and the bispectrum, therefore the signal to noise increases for both observables.
For the power spectrum, this translates to proportionally tighter constraints on the parameters, compared to the fiducial model.
For the bispectrum the picture is more complicated:
conditional error bars shrink as expected; 
but correlations among different parameters become more severe, enlarging the marginalized error bars.
In any case, the marginalized errors obtained with the bispectrum, even with this model, still improve the PS one by approximately one order of magnitude.
Therefore, our conclusions are qualitatively unchanged.
Even more than in the fiducial case, combining power spectrum and bispectrum improves the constraining power and reliability of the results.
Since the correlations are particularly severe for a subset of parameters, fixing even a limited number of them would allow for recovering most of the constraining power we forecasted with the Tinker fiducial model, as seen in figure \ref{fig:NoNoiseNoForeground5000_Tinker_zMCut}.
Nevertheless we conservatively quote as marginalized errors the results obtained without fixing any parameter.

\begin{table}
    \centering
    \begin{tabular}{|c|ccc|ccc|}
    \hline
    & \multicolumn{3}{c|}{Conditional} & \multicolumn{3}{c|}{Marginalized}\\
                & PS $1\sigma$ & BS $1\sigma$ & PS$\oplus$BS $1\sigma$ & PS $1\sigma$ & BS $1\sigma$ & PS$\oplus$BS $1\sigma$\\
    \hline
$\Omega_m$	&	0.00041	&	0.0016	&	0.00030	&	0.59	&	0.11	&	0.037	\\
$h$	&	0.0038	&	0.0072	&	0.0033	&	13	&	0.60	&	0.49	\\
$\sigma_8$	&	0.00038	&	0.0013	&	0.00029	&	0.35	&	0.065	&	0.035	\\
$n_S$	&	0.0046	&	0.022	&	0.0030	&	11	&	0.50	&	0.39	\\
$w_0$	&	0.0048	&	0.015	&	0.0040	&	5.3	&	0.51	&	0.25	\\
$b_\text{HSM}$	&	0.0012	&	0.0039	&	0.00089	&	5.6	&	0.89	&	0.77	\\
$P_0$	&	0.017	&	0.056	&	0.013	&	147	&	13	&	9.1	\\
$\alpha_P$	&	0.0046	&	0.021	&	0.0031	&	2.5	&	0.15	&	0.13	\\
$c_{500,c}$	&	0.0013	&	0.0049	&	0.00092	&	0.33	&	0.12	&	0.10	\\
$\alpha_G$	&	0.00078	&	0.0024	&	0.00057	&	2.5	&	0.086	&	0.078	\\
$\beta_G$	&	0.0058	&	0.017	&	0.0038	&	2.2	&	0.46	&	0.37	\\
$\gamma_G$	&	0.0016	&	0.0043	&	0.0013	&	4.0	&	0.15	&	0.11	\\
    \hline
    \end{tabular}
    \caption{
    Same as table \ref{tab:results_noiseless_ell5000} (CVL, $\ell_\text{max}=5000$,  $f_\text{sky}=1$) but with the halo redshift and mass boundaries observationally set by \Planck.}
    \label{tab:results_noiseless_ell5000_zMcut}
\end{table}


\subsubsection{Analysis of a subset of configurations}
\label{sec:EqSqAnalyis}

To validate our statement that the bispectrum is effective at breaking degeneracies because of the shape-dependent response to changes in parameters, combined with the large number of available triangles, we repeat our analysis limiting ourselves to only equilateral configurations, only squeezed configurations, and a combination of the two.

If we consider conditional errors (left part of table \ref{tab:results_EqSq}), the power spectrum constraining power exceeds the bispectrum
The equilateral configurations contribute the most for the bispectrum, whereas the squeezed ones have one order of magnitude less constraining power.
Upon marginalization (right part of the table), the constraining power of the squeezed and equilateral bispectrum degrades, as it does for the power spectrum,
with the squeezed one degrading noticeably less.
However, when we jointly analyze equilateral and squeezed limits, the degeneracies start to break and the bispectrum marginalized errors start to be competitive against the power spectrum ones.
Obviously, the full bispectrum comprises even more kinds of configuration that can further ease the remaining degeneracies.
We can therefore conclude that, as we claimed in section~\ref{sec:Derivatives}, the reason why the bispectrum constraining power is not as sensitive as the power spectrum one to parameter degeneracies can be explained with the presence of many more modes that are all weighted differently when the value of the parameters changes.

\begin{table}
    \centering
    \begin{tabular}{|c|cccc|cccc|}
    \hline
    & \multicolumn{4}{c|}{Conditional} & \multicolumn{4}{c|}{Marginalized}\\
                & $\frac{\sigma^\text{Eq}}{\sigma^\text{PS}}$
                & $\frac{\sigma^\text{Sq}}{\sigma^\text{PS}}$
                & $\frac{\sigma^\text{Eq+Sq}}{\sigma^\text{PS}}$
                & $\frac{\sigma^\text{Full}}{\sigma^\text{PS}}$
                & $\frac{\sigma^\text{Eq}}{\sigma^\text{PS}}$
                & $\frac{\sigma^\text{Sq}}{\sigma^\text{PS}}$
                & $\frac{\sigma^\text{Eq+Sq}}{\sigma^\text{PS}}$
                & $\frac{\sigma^\text{Full}}{\sigma^\text{PS}}$\\[2pt]
    \hline
$\Omega_m$	&	5.9	&	86	&	5.9	&	4.0	&	10	&	5.7	&	0.65	&	0.029	\\
$h$	&	2.6	&	29	&	2.6	&	2.0	&	13	&	23	&	0.99	&	0.015	\\
$\sigma_8$	&	4.7	&	74	&	4.7	&	3.3	&	11	&	25	&	3.2	&	0.025	\\
$n_S$	&	6.1	&	41	&	6.0	&	5.1	&	43	&	217	&	7.4	&	0.10	\\
$w_0$	&	4.5	&	184	&	4.5	&	2.8	&	3.3	&	14	&	0.67	&	0.0094	\\
$b_\text{HSM}$	&	4.9	&	71	&	4.9	&	3.5	&	0.98	&	2.0	&	0.11	&	0.0021	\\
$P_0$	&	4.7	&	75	&	4.7	&	3.3	&	1.2	&	1.5	&	0.12	&	0.0015	\\
$\alpha_P$	&	6.0	&	65	&	6.0	&	4.3	&	3.7	&	6.0	&	0.55	&	0.014	\\
$c_{500,c}$	&	5.9	&	70	&	5.9	&	3.9	&	8.4	&	11	&	2.0	&	0.018	\\
$\alpha_G$	&	5.2	&	81	&	5.2	&	3.1	&	1.1	&	2.3	&	0.21	&	0.0019	\\
$\beta_G$	&	6.7	&	57	&	6.7	&	1.1	&	3.4	&	37	&	1.4	&	0.025	\\
$\gamma_G$	&	3.9	&	73	&	3.9	&	2.7	&	1.9	&	0.85	&	0.085	&	0.00079	\\
    \hline
    \end{tabular}
    \caption{Ratio of the equilateral (Eq.), squeezed (Sq.), joint equilateral-plus-squeezed (Eq.+Sq.), and full bispectrum error to the power spectrum one.
    After marginalization, the constraining power of the bispectrum in one single limit is degraded comparably to the power-spectrum.
    The same does not apply when we consider the combination of equilateral and squeezed limit.
    Obviously, the full bispectrum --- that encompasses many other configurations, orthogonals and all the intermediate ones --- is even more stable under the marginalization.}
    \label{tab:results_EqSq}
\end{table}

\subsection{Validation against Planck noise and foreground contamination}
The \textit{Planck} satellite has provided a full sky map of the Compton-$y$ parameter \cite{Aghanim:2015eva}.
This data was used to measure the the tSZ power spectrum and the bispectrum, and the former was used to constraint cosmological parameters.
It is interesting to understand what are the prospective results for a full fledged analysis of the bispectrum recovered from \textit{Planck} data.
To do so, we consider the \textit{Planck} Gaussian noise and foreground determined in \cite{Aghanim:2015eva}, and 
we restrict our analysis in the multipole range $[70, 1000]$ to remove the bulk of non-Gaussian noise and remaining spurious contributions.
This was performed according to what has been done in \cite{Aghanim:2015eva, Lacasa:2014gha}.
We show our results in table \ref{tab:results_ell1000_PlanckNoise_FiducialTinker}.
As it is well known, the parameters cannot be meaningfully constrained separately,
since are degenerate at power spectrum level.
Insead, they can be combined in a parameter that controls the overall amplitude, such as $F=\sigma_8 \, \Omega_m^{0.4} (1-b)^{-0.4} h^{-0.21}$ \cite{Bolliet:2017lha}.
In fact, we see that after marginalization the relative errors are $\mathcal{O}(100)$, while using the full bispectrum even \textit{Planck} data could lead to $\mathcal{O}(1)$ relative errors even after marginalizing over all other parameters.

\begin{table}
    \centering
    \begin{tabular}{|c|ccc|ccc|}
    \hline
    & \multicolumn{3}{c|}{Conditional} & \multicolumn{3}{c|}{Marginalized}\\
                & PS $1\sigma$ & BS $1\sigma$ & PS$\oplus$BS $1\sigma$ & PS $1\sigma$ & BS $1\sigma$ & PS$\oplus$BS $1\sigma$\\
    \hline
$\Omega_m$	& 0.0036	& 0.015	& 0.0026    & 13	    & 0.26	& 0.19	\\ 
$h	$	    & 0.022	    & 0.054	& 0.017	    & 24	    & 0.94	& 0.86	\\ 
$\sigma_8$	& 0.0035	& 0.012	& 0.0026	& 5.9	    & 0.12	& 0.12	\\ 
$n_S	$	& 0.092	    & 0.098	& 0.079	    & 13	    & 1.0 	& 0.77	\\ 
$w_0	$	& 0.19	    & 0.62	& 0.15	    & 160   	& 1.4	& 1.3	\\ 
$b_\text{HSM}$	& 0.01	& 0.038	& 0.0078	& 610   	& 2.0 	& 1.8	\\ 
$P_0	$	& 0.17	    & 0.59	& 0.13	    & 12000	    & 33	& 29	\\ 
$\alpha_P$	& 0.19	    & 0.12	& 0.12	    & 100	    & 0.39	& 0.37	\\ 
$c_{500,c}$	& 0.01	    & 0.041	& 0.0074	& 7.0	    & 0.18	& 0.18	\\ 
$\alpha_G$	& 0.007	    & 0.025	& 0.0053	& 100	    & 0.17	& 0.17	\\ 
$\beta_G$	& 0.039	    & 0.077	& 0.027	    & 3.1	    & 0.092	& 0.092	\\ 
$\gamma_G$	& 0.02	    & 0.057	& 0.015	    & 370   	& 0.54	& 0.48	\\ 
    \hline
    \end{tabular}
    \caption{Conditional and marginalized error forecasts for an experiment with \Planck-like noise level and foregrounds, $\ell_\text{max}=1000$, and $f_\text{sky}=0.47$.}
    \label{tab:results_ell1000_PlanckNoise_FiducialTinker}
\end{table}


Introducing the cuts in mass and redshift have a comparatively similar impact as in the previous case, as shown in table \ref{tab:results_ell1000_PlanckNoise_zMCuts}.

\begin{table}
    \centering
    \begin{tabular}{|c|ccc|ccc|}
    \hline
    & \multicolumn{3}{c|}{Conditional} & \multicolumn{3}{c|}{Marginalized}\\
                & PS $1\sigma$ & BS $1\sigma$ & PS$\oplus$BS $1\sigma$ & PS $1\sigma$ & BS $1\sigma$ & PS$\oplus$BS $1\sigma$\\
    \hline
$\Omega_m$	& 0.0027	& 0.0078	& 0.0018	& 2.4	& 1.4	& 0.18	\\ 
$h	$	    & 0.017     & 0.032 	& 0.015	    & 51	& 4.3	& 4.1	\\ 
$\sigma_8$	& 0.0027	& 0.0072	& 0.0019	& 2.6	& 0.53	& 0.25	\\ 
$n_S	$	& 0.13	    & 0.076 	& 0.07	    & 48	& 6.3	& 3.2	\\ 
$w_0	$	& 0.14	    & 0.29	    & 0.12	    & 31	& 4.4	& 1.5	\\ 
$b_\text{HSM}$	& 0.0076	& 0.018	    & 0.0056	& 55	& 6.3	& 5.2	\\ 
$P_0	$	& 0.12	    & 0.29	    & 0.095	    & 2100	& 120	& 83	\\ 
$\alpha_P$	& 0.24	    & 0.073	    & 0.051	    & 24	& 1.4	& 1.1	\\ 
$c_{500,c}$	& 0.0074	& 0.02	    & 0.0052	& 2.7	& 0.83	& 0.76	\\ 
$\alpha_G$	& 0.0052	& 0.012	    & 0.0039	& 32	& 0.72	& 0.64	\\ 
$\beta_G$	& 0.028	    & 0.086	    & 0.019	    & 17	& 3.4	& 2.7	\\ 
$\gamma_G$	& 0.015	    & 0.03	    & 0.012	    & 58	& 1.5	& 1.0	\\ 
    \hline
    \end{tabular}
    \caption{
    Same as table \ref{tab:results_ell1000_PlanckNoise_FiducialTinker} (\Planck-like noise and foregrounds, $\ell_\text{max}=1000$, $f_\text{sky}=0.47$) but with the halo redshift and mass boundaries observationally set by \Planck.}
    \label{tab:results_ell1000_PlanckNoise_zMCuts}
\end{table}


\subsection{Forecast for future realistic surveys}
To assess what are the prospects of full bispectrum analysis in the near future and in the mid-term, we use the results on section \ref{sec:GaussianForegrounds} in the Fisher matrix formalism to model the impact of Gaussian noise and foreground.
As already discussed, the bulk of non-Gaussian contaminations can be removed enforcing a cut on the lowest multipole, which has little to no impact on the forecasted constraints.

In particular we consider two specific surveys, that are representative of what can be achieved in the next decade and in the next $\approx 30$ years: Simons Observatory and the proposed Voyage 2050 Spectro-Polarimeter, respectively.

Simons Observatory (SO) \cite{Ade:2018sbj}  is an observational facility currently being built in the Atacama desert, that will be devoted to the measurement of gravitational lensing, tSZ effect, CMB temperature and polarization on very small scales.
It comprises one 6 m Large Aperture Telescope (LAT) and three 0.5 m Small Aperture Telescopes.
Here we just consider the LAT instrument, whose noise profile is stated in table 1 of \cite{Ade:2018sbj}.
The observation field will amount to 40\% of the sky but, since we expect that some masking will be anyway needed, we conservatively set $f_\text{sky} = 0.30$ in the relative forecast.
The results are reported in table~\ref{tab:results_SONoise}.
Despite the lower sky coverage, the higher raw sensitivity and beam size will allow SO to greatly improve constraints over \textit{Planck}.

\begin{table}
    \centering
    \begin{tabular}{|c|ccc|ccc|}
    \hline
    & \multicolumn{3}{c|}{Conditional} & \multicolumn{3}{c|}{Marginalized}\\
                & PS $1\sigma$ & BS $1\sigma$ & PS$\oplus$BS $1\sigma$ & PS $1\sigma$ & BS $1\sigma$ & PS$\oplus$BS $1\sigma$\\
    \hline
$\Omega_m$	& 0.00081	& 0.0033	& 0.00058	& 1.8	& 0.057	& 0.04	\\ 
$h	$	    & 0.0074	& 0.015	    & 0.0062	& 12	& 0.18	& 0.17	\\ 
$\sigma_8$	& 0.00075	& 0.0025	& 0.00056	& 0.91	& 0.024	& 0.023	\\ 
$n_S	$	& 0.0095	& 0.049	    & 0.0057	& 1.6	& 0.18	& 0.14	\\ 
$w_0	$	& 0.0097	& 0.027	    & 0.0075	& 23	& 0.22	& 0.20	\\ 
$b_\text{HSM}$	& 0.0023	& 0.0082	& 0.0017	& 160	& 0.37	& 0.35	\\ 
$P_0	$	& 0.034	    & 0.11	    & 0.025	    & 3700	& 6.0	    & 5.6	\\ 
$\alpha_P$	& 0.0096	& 0.041	    & 0.0058	& 5.4	& 0.089	& 0.083	\\ 
$c_{500,c}$	& 0.0026	& 0.011	    & 0.0018	& 1.9	& 0.035	& 0.034	\\ 
$\alpha_G$	& 0.0015	& 0.0049	& 0.0011	& 16	& 0.032	& 0.031	\\ 
$\beta_G$	& 0.012	    & 0.015	    & 0.007	    & 0.66	& 0.017	& 0.017	\\ 
$\gamma_G$	& 0.0032	& 0.0086	& 0.0025	& 100	& 0.085	& 0.079	\\ 
    \hline
    \end{tabular}
    \caption{
    Conditional and marginalized error forecasts for an experiment with SO-like noise level and foregrounds, $\ell_\text{max}=5000$, and $f_\text{sky}=0.30$.}
    \label{tab:results_SONoise}
\end{table}


The Voyage 2050 Spectro-Polarimeter (V-SP) \cite{Delabrouille:2019thj} is an L-class mission concept that has been put forward for the ESA call Voyage 2050.
V-SP will map the whole sky, and therefore we conservatively assume it will use a mask much similar to the one employed by Planck to generate the $y$ map.
Hence, when forecasting its performance we will take $f_\text{sky}=0.47$.
We use the noise profile reported in table 1 of \cite{Delabrouille:2019thj}.
Table \ref{tab:results_SPADENoise} show the results in this scenario; the performance incrementally increases over the SO forecast.

\begin{table}
    \centering
    \begin{tabular}{|c|ccc|ccc|}
    \hline
    & \multicolumn{3}{c|}{Conditional} & \multicolumn{3}{c|}{Marginalized}\\
                & PS $1\sigma$ & BS $1\sigma$ & PS$\oplus$BS $1\sigma$ & PS $1\sigma$ & BS $1\sigma$ & PS$\oplus$BS $1\sigma$\\
    \hline
$\Omega_m$	& 0.00064	& 0.0025	& 0.00045	& 1.3	& 0.037	& 0.026	\\ 
$h	$	    & 0.0059	& 0.012	    & 0.0049	& 8.3	& 0.12	& 0.11	\\ 
$\sigma_8$	& 0.00059	& 0.0019	& 0.00043	& 0.63	& 0.016	& 0.015	\\ 
$n_S	$	& 0.0073	& 0.037	    & 0.0043	& 1.1	& 0.12	& 0.093	\\ 
$w_0	$	& 0.0076	& 0.021	    & 0.0057	& 16	& 0.15	& 0.14	\\ 
$b_\text{HSM}$	& 0.0018	& 0.0064	& 0.0013	& 110	& 0.24	& 0.23	\\ 
$P_0	$	& 0.027	    & 0.088	    & 0.02	    & 2600	& 4.0	& 3.7	\\ 
$\alpha_P$	& 0.0074	& 0.032	    & 0.0044	& 4.1	& 0.059	& 0.054	\\ 
$c_{500,c}$	& 0.002	    & 0.0079	& 0.0014	& 1.3	& 0.023	& 0.022	\\ 
$\alpha_G$	& 0.0012	& 0.0037	& 0.00086	& 11	& 0.022	& 0.021	\\ 
$\beta_G$	& 0.0091	& 0.0099	& 0.0053	& 0.46	& 0.011	& 0.011	\\ 
$\gamma_G$	& 0.0025	& 0.0066	& 0.0019	& 73	& 0.058	& 0.054	\\ 
    \hline
    \end{tabular}
    \caption{
    Conditional and marginalized error forecasts for an experiment with V-SP-like noise level and foregrounds, $\ell_\text{max}=5000$, and $f_\text{sky}=0.47$.}
    \label{tab:results_SPADENoise}
\end{table}


It is interesting to notice that V-SP will already be close to saturating the cosmic variance limited constraints.
Since from the technological point of view we could already have the raw sensitivity to get close to the cosmic variance, this reinforces the need for developing new better methods of foreground removal, that will allow us to make use of this potential.

For a visual comparison between a CVL survey, \textit{Planck}, SO and V-SP, we refer to appendix \ref{sec:A_FullTrianglePlot}.

\section{Discussion and Conclusions}\label{sec:Conclusions}

In this work, we have carried out a joint Fisher analysis of the thermal Sunyaev-Zeldovich power spectrum and bispectrum. 
In our analysis, we have gone beyond similar studies by significantly increasing the level of accuracy and realism of our forecasts. This has been done in several ways. First of all, we have evaluated the full power spectrum and bispectrum (auto- and cross-) covariance in the analysis, including all NG contributions. This turns out to be important, since we have found that the covariance is dominated by the connected 6-point component. Furthermore, we have considered a multi-component energy spectrum scenario and we have modeled the effects of component separation in our forecasts, via an effective noise term, which was obtained from a simple ILC procedure. 
Finally, rather than focusing just on $\sigma_8$ or on a small number of parameters, we have considered an extended set of both cosmological and ICM parameters, with the aim to accurately assess correlations and degeneracies and how these are dealt with by both the power spectrum and bispectrum.

In the end, we find out that the tSZ bispectrum is a very powerful observable, able to produce even stronger constraints than the tSZ power spectrum, after marginalization in a multi-parameter analysis. This is shown in our main results, summarized in figure \ref{fig:ErrorBars_NoNoiseNoForeground5000} and table~\ref{tab:results_noiseless_ell5000_zMcut}. 

Several reasons have been already pointed out in previous studies, which explain why the bispectrum is so useful in this type of analysis. For example it has been observed that the bispectrum is less affected by uncertainties in ICM parameters, because its contributions come from (better understood) low-redshift, high-mass clusters \cite{2012ApJ...760....5B}. Another important point is that the bispectrum can break degeneracies that are present in a power spectrum-only analysis, through a different amplitude scaling with parameters \cite{Hill:2012ec, Hurier:2017jgi}.

On top of these previously known aspects, our main finding is that the bispectrum is extremely efficient at breaking degeneracies not only via amplitude scaling, but also --- and mostly --- due to the fact that different triangle {\em shapes} (e.g. equilateral and squeezed triangles) are affected differently by parameter changes. This is a somewhat counter-intuitive result: tSZ statistics are dominated by the one-halo term, which leads to a weak shape dependence on cosmology.
Yet, we see that this is already strong enough in the bispectrum to produce large improvements in the final results. Furthermore, the triangle shape dependence on ICM parameters is of course stronger than on cosmology, significantly lowering the impact of astrophysical uncertainties in the cosmological analysis (as well as allowing for a precise measurement of astrophysical parameters themselves, possibly complementing \cite{Pandey:2019uxy}).
Given its importance, we have studied this effect in detail. To this purpose, we have isolated two specific types of triangles, namely equilateral and squeezed. We have then verified that including only one configuration type in the bispectrum analysis leads to only modest improvements in the final results, whereas including both at the same time produces significantly better forecasts; we have checked that is precisely due to a slightly different response to changes in parameters, in the squeezed and equilateral limit.
To further investigate such behaviour, we have isolated the regions in the $M$-$z$ plane which mostly contribute to the bispectrum in different limits, showing that such regions do not fully overlap. In other words, when we compute equilateral and squeezed bispectrum configurations, we effectively integrate the halo mass function over slightly different intervals in mass and redshift. The effect of this on parameters is fairly small for a single triangle. However, this adds up in a significant way when we produce the final forecasts, by integrating over the very large number of available configurations.

The results obtained in our study clearly suggest that a joint power spectrum-bispectrum analysis of, e.g., {\it Planck} data, or a complete likelihood study of this kind, using mock datasets for future experiments, is clearly worth pursuing. This is the object of ongoing work. Another interesting subject for future investigation consists in accounting for spectral corrections to tSZ distortions, arising from relativistic speeds of electrons in clusters \cite{Chluba:2012dv} (the so called relativistic SZ). Such corrections are mass and redshift dependent \cite{Remazeilles:2018laq, Remazeilles:2019mld}
through the cluster temperature scaling \cite{Lee:2019tes} and therefore could provide further help to break parameter degeneracies if better sensitivity is achieved.
Further synergies, motivated by the recent interest in the bispectrum of other cosmological probes \cite{Yankelevich_2018,2019MNRAS.490.4688R, Hahn:2019zob, Karagiannis:2018jdt, Hill:2018ypf,Heinrich:2020xub}, could be achieved considering the cross-bispectrum of tSZ with other tracers of the low-redshift Universe matter distribution.

\acknowledgments
We thank Boris Bolliet and Jens Chluba for useful discussion, and Aditya Rotti for extensive comments on an early version of the draft. 
AR was supported by the ERC Consolidator Grant {\it CMBSPEC} (No.~725456) as part of the European Union's Horizon 2020 research and innovation program. 
ML was supported by the project ``Combining Cosmic Microwave Background and Large Scale Structure data: an Integrated Approach for Addressing Fundamental Questions in Cosmology'', funded by the MIUR Progetti di Ricerca di Rilevante Interesse Nazionale (PRIN) Bando 2017 - grant 2017YJYZAH. ML also acknowledges support by the University of Padova under the STARS Grants programme CoGITO, Cosmology beyond Gaussianity, Inference, Theory and Observations.
FL was supported by a postdoctoral grant from Centre National d’\'{E}tudes Spatiales (CNES).
This work has made use of the Horizon Cluster hosted by Institut d'Astrophysique de Paris. We thank Stephane Rouberol for maintenance and running of this computing cluster.

\newpage

\appendix

\section{Projection of the 3D tSZ field}
\label{sec:A_ProjectedAngularCorrelators}

\subsection{Correlation functions for the projected Compton-%
\texorpdfstring{$y$}{y}
 field}
The total Compton-$y$ field in a given direction $\myvec{n}$ of the sky is defined via the following line-of-sight integral \citep{Hill:2013baa}
\begin{equation} 
\label{2Dy}
    y\left(\myvec{n}\right) = \int \diff t\ y_{3\mathrm{D}}\left(\chi(t)\myvec{n}\right) = \int \diff \chi\ a\left(\chi\right) y_{3\mathrm{D}}\left(\chi(t)\myvec{n}\right) \, .
\end{equation}
The three-dimensional Compton-$y$ field is directly related to the electron pressure profile of a single halo $P_{\mathrm{e}}$ via the following re-scaling
\begin{equation}
    \label{3Dy}
    y_{3\mathrm{D}}\left(\myvec{r}\right) = \frac{\sigma_{\mathrm{T}}}{m_{\mathrm{e}}c^2}P_{\mathrm{e}}\left(\myvec{r}\right),
\end{equation}
$\myvec{r}$ being the comoving distance from the center of the halo. Under the assumption of small enough scales, we can focus on the two-dimensional Fourier transform of the field \eqref{2Dy} employing the flat-sky approximation 
\begin{equation}
    \label{need20}
    \tilde{y}_{\boldsymbol{\ell}} = \int d\boldsymbol{\theta}_{\mathrm{n}}\ y\left(\boldsymbol{\theta}_{\mathrm{n}}\right)
    e^{i\boldsymbol{\ell}\cdot\boldsymbol{\theta}_{\mathrm{n}}} \, ,
    \qquad
    y\left(\boldsymbol{\theta}_{\mathrm{n}}\right)\equiv \int \diff \chi\ a\left(\chi\right) y_{3\mathrm{D}}\left(\chi,\boldsymbol{\theta}_{\mathrm{n}}\chi\right).
\end{equation}
In Fourier space, we define the connected part of the $n$-point correlation function of the two-dimensional field $\tilde{y}_{\boldsymbol{\ell}}$ 
\begin{equation}
\label{CorrFunct}
\langle\tilde{y}_{\boldsymbol{\ell}_1} \dots \tilde{y}_{\boldsymbol{\ell}_2}\rangle_c \equiv \left(2\pi\right)^2 P_{n}\left(\boldsymbol\ell_1, \dots,\boldsymbol\ell_n\right) \delta_{\mathrm{D}}\left(\boldsymbol\ell_1 + \dots + \boldsymbol\ell_n\right).
\end{equation}
 We call the quantity $P_{n}\left(\boldsymbol\ell_1, \dots,\boldsymbol\ell_n\right)$ polyspectrum of order $n$. In eq.~\eqref{CorrFunct} we described the field on a flat sky, i.e. we approximate the full spherical harmonics decomposition of the real field with a simple two-dimensional Fourier transform. By replacing eq. \eqref{need20} into eq.~\eqref{CorrFunct}, we can derive the expression for the flat-sky polyspectra. In general, the redshift integration appearing in eq.~\eqref{need20} would naturally translate into a complex $n$-dimensional one. To simplify this calculation, we make use of the Limber approximation \citep{2008PhRvD..78l3506L}: we assume that the three-dimensional matter polyspectra have a weak dependence on the momenta component corresponding to the line-of-sight direction. 
Consequently, the projection collapses into a simple one-dimension redshift integration and we can relate the angular multipoles $\boldsymbol{\ell}$ to the three-dimensional momenta $\mathbf{k}$ via the well known Limber relation $\mathbf{k}\left(\Bell,z\right)\approx \{\Bell / \chi\left(z\right),0\}$. Finally, the general $n$-point polyspectrum $P_{n} \left(\boldsymbol{\ell}_1, \dots,\boldsymbol{\ell}_n\right)$ relates to the same order one  $P^{y_{3\mathrm{D}}}_n\left(\mathbf{k}_1, \dots,\mathbf{k}_n\right)$ for the three-dimensional Compton-$y$ field \eqref{3Dy} via 
\begin{equation}
\begin{split}
    P_{n}\left(\boldsymbol\ell_1, \dots,\boldsymbol\ell_n\right)  
    =
    &
    \int _0^{\infty} \mathrm{d}\chi\ \chi^{2-2n}\ a^n\left(\chi\right)  P^{y_{3\mathrm{D}}}_n\left(\mathbf{k}\left(\Bell_1,\chi\right), \dots,\mathbf{k}\left(\Bell_n,\chi\right)\right) 
\\
    =
    &
    \int _0^{\infty} \mathrm{d}z\ \mathcal{Q}^{(n)}(z)  P^{y_{3\mathrm{D}}}_n\left(\mathbf{k}\left(\Bell_1,z\right), \dots,\mathbf{k}\left(\Bell_n,\chi\right)\right) \, ,
\label{nProjection}
\end{split}
\end{equation}
where we define the kernel 
\begin{equation}
   \mathcal{Q}^{(n)}(z)\equiv \chi^{2-2n}(z)\ a^n\left(z\right) \frac{\diff \chi}{\diff z}  = \frac{\chi^{2-2n}(z)}{H(z)}a^n\left(z\right).
\end{equation}
For sake of completeness, let us recall the definition for the quantity $P^{y_{3\mathrm{D}}}_n\left(\myvec{k}_1, \dots,\myvec{k}_n\right)$
\begin{align}
\label{3Dpoly}
   \langle\tilde{y}_{3\mathrm{D}}\left(\myvec{k}_1\right)\dots\tilde{y}_{3\mathrm{D}}\left(\myvec{k}_n\right)\rangle_c 
   \equiv
   &
   \left(2\pi\right)^3 P^{y_{3\mathrm{D}}}_n\left(\myvec{k}_1, \dots,\myvec{k}_n\right) \delta_{\mathrm{D}}\left(\myvec{k}_1 + \dots + \myvec{k}_n\right),
\\
   \tilde{y}_{3\mathrm{D}}\left(\myvec{k}\right)
   =
   &
   \int \diff^3 \myvec{x}\ y_{3\mathrm{D}}\left(\myvec{x}\right) e^{+i\myvec{k}\cdot\myvec{x}}.
\end{align}
As we do rely on the halo model for the matter clustering, we can write the electron pressure profile $y_{3\mathrm{D}}\left(\myvec{x}\right)\propto P_{\mathrm{e}}\left(\myvec{x}\right)$ \eqref{3Dy} (and its Fourier transform) as the sum of contributions from different halos, the mass of the $i^{\mathrm{th}}$ halo being $m_i$, centered at position $\myvec{x}_i$ 
\begin{equation}
\label{y3dtothalo}
\tilde{y}_{3\mathrm{D}}\left(\myvec{k}\right)  = \sum_{i}^{n^{\circ}\ \mathrm{halos}} \int \diff m \diff^3\myvec{x}\      \tilde{y}_{3\mathrm{D}}\left(\myvec{k},m\right)e^{+i\myvec{x}\cdot\myvec{k}}\ \delta\left(m-m_i\right) \delta\left(\myvec{x} - \myvec{x}_i\right).
\end{equation}
The contribution $\tilde{y}_{3\mathrm{D}}\left(\myvec{k},m\right)$ from the single halo of mass $m^{\mathrm{c}}_{500}$ can be obtained as  
\begin{equation}
\begin{split}
\label{y3d1halo}
    \tilde{y}_{3\mathrm{D}}\left(\myvec{k},m\right)
    =
    & \ 
    4\pi\frac{\sigma_{\mathrm{T}}}{m_{\mathrm{e}}c^2}\int \diff r\ r^2 j_0\left(kr\right)P_{\mathrm{e}}\left(m,r\right)
\\
    =
    & \ 
    4\pi r_{500}\frac{\sigma_{\mathrm{T}}}{m_{\mathrm{e}}c^2}\int \diff x\ x^2 j_0\left(kxr_{500}\right)P_{\mathrm{e}}\left(m^{\mathrm{c}}_{500},x\right) \, ,
\end{split}
\end{equation}
where we employed the mass definition $m_{500}^{\mathrm{c}}$ and the variable $x$ as requested for employing the parametrisation $\eqref{Pe}$.
By replacing eq. \eqref{y3dtothalo} within eq. \eqref{3Dpoly}, it is possible to obtain any desired order of correlation for the three-dimensional Compton-$y$ parameter. Furthermore, by splitting the sum in eq. \eqref{y3dtothalo} into contributions from different multi-halo configurations, we do obtain the well known hierarchy of the different halo terms. We can finally obtain the observable of interest via eq. \eqref{CorrFunct}. 
For consistency with the literature, we call the 2-, the 3- and the 4-point polyspectrum power spectrum, bispectrum and trispectrum, respectively:
\begin{align}
C_{\ell} & \equiv P_{2}\left(\Bell\right),\label{RenamePS}\\
b_{\ell_1\ell_2\ell_3} &\equiv P_{3}\left(\Bell_1,\Bell_2,\Bell_3\right),\label{RenameBS}\\
T\left(\Bell_1,\Bell_2,\Bell_3,\Bell_4\right) &\equiv P_{4}\left(\Bell_1,\Bell_2,\Bell_3,\Bell_4\right).\label{RenameT}
\end{align}
We underline that the assumption of an isotropic and homogeneous Universe (Cosmological Principle) allows us to reduce the actual dependencies of the polyspectra. The power spectrum is expressed as function of the module $\ell$ of the momentum $\Bell$ (we define $\ell \equiv |\Bell|$) and the bispectrum has a dependence on just 3 degrees of freedom, i.e. the edges of the associated triangular configuration \citep{2017JCAP...02..032M}. In the following, we will show in details the equations required for our implementation as obtained from the formalism above.

\subsection{Data vector}
As far as the data vector of our analyses is concerned, we can write the power spectrum as the sum of the one- and two-halo term
\begin{align}
\label{PspHMtot}
    C_{\ell} 
    =
    & \ 
    C^{1\mathrm{h}}_{\ell} + C^{2\mathrm{h}}_{\ell} \, ,
\\
    C^{1\mathrm{h}}_{\ell}
    =
    & 
    \int \diff z\ \mathcal{Q}^{(2)}(z) \int \diff\nu\ f\left(\nu,z\right)\frac{m_{\nu}}{\rho_{\mathrm{m},0}}\ \big{|}\tilde{y}_{3\mathrm{D}}\left(k_{\ell}^z,m_{\nu}\right)\big{|}^2 \, ,
\\
    C^{2\mathrm{h}}_{\ell}\left(\ell\right) 
    =
    &
    \int \diff z \ \mathcal{Q}^{(2)}(z)\ P^{\mathrm{lin.}}(k^z_{\ell},z) \bigg[\int \diff\nu\ f\left(\nu,z\right)\frac{m_{\nu}}{\rho_{\mathrm{m},0}}\  b^{(1)}(z, m_{\nu})\  \tilde{y}_{3\mathrm{D}}\left(k_{\ell}^z,m_{\nu}\right) \bigg]^2 \, ,
\end{align}
where we introduced the abbreviation $k^z_{\ell}\equiv |\myvec{k}\left(\Bell,\chi\left(z\right)\right)|$ and the mass $m_{\nu}$ is related to the parameter $\nu$ via $\sigma(m_{\nu})=\delta/\nu$.
To simplify the expression for the general matter polyspectrum, we can introduce the following quantity
\begin{equation}
\label{ImunuHalos}
\text{I}_{\mu}^{\beta}\left(k_1, \dots, k_{\mu};z\right) = \int \diff\nu\ f\left(\nu,z\right)\frac{m_{\nu}}{\rho_{\mathrm{m},0}} \ b^{(\beta)}(z,m_{\nu})\ \prod_{i=1}^{\mu} \tilde{y}_{3\mathrm{D}}\left(k_i,m_{\nu}\right)  ,
\end{equation}
where $b^{(0)} \equiv 1$ and we omitted the redshift dependence.
Then, the Compton-$y$ power spectrum~\eqref{PspHMtot} can be written in a more synthetic way as 
\begin{equation}
\label{PspHMtot2}C_{\ell} = \int \diff z \ \mathcal{Q}^{(2)}(z) \left[ \text{I}_{2}^{0}\left(k^z_{\ell},k^z_{\ell},z\right) + \left[\text{I}_{1}^{1}\left(k^z_{\ell},z\right)\right]^2 P^{\text{lin.}}\left(k^z_{\ell},z\right) \right].
\end{equation}
Borrowing the above notation, we can write the bispectrum used in this work as the sum of the one- and two-halo term 
\begin{align}
\label{totBisp}
    b_{\ell_1\ell_2\ell_3} 
    =
    & \ 
    b^{1\mathrm{h}}_{\ell_1\ell_2\ell_3} + 
     b^{2\mathrm{h}}_{\ell_1\ell_2\ell_3} \, ,
\\
    b^{1\mathrm{h}}_{\ell_1\ell_2\ell_3} 
    =
    &
    \int \diff z \ \mathcal{Q}^{(3)}(z)\ \text{I}_{3}^{0}
    \left(k^z_{\ell_1},k^z_{\ell_2},k^z_{\ell_3};z\right) ,
\\
    \begin{split}
        b^{2\mathrm{h}}_{\ell_1\ell_2\ell_3} 
        =
        & 
        \int \diff z \ \mathcal{Q}^{(3)}(z)\ \Big{(} \text{I}_1^1\left(k^z_{\ell_1}\right)\text{I}_2^1\left(k^z_{\ell_2},k^z_{\ell_3}\right)P^{\text{lin.}}\left(k^z_{\ell_1},z\right)
    \\ 
        &\quad+ 				 \text{I}_1^1\left(k^z_{\ell_3}\right)\text{I}_2^1\left(k^z_{\ell_1},k^z_{\ell_2}\right)P^{\text{lin.}}\left(k^z_{\ell_3},z\right) +
        \text{I}_1^1\left(k^z_{\ell_2}\right)\text{I}_2^1\left(k^z_{\ell_3},k^z_{\ell_1}\right)P^{\text{lin.}}\left(k^z_{\ell_2},z\right)\Big{)}.
    \end{split}
\end{align}
As already mentioned in the main text, we do not include the three-halo term as it would require the second order halo bias for which a fit has not yet been performed.
\subsection{Covariance matrix}
Moving to the covariance matrix, we employ the full expressions \eqref{PspHMtot} and \eqref{totBisp} whenever power spectra and bispectra are required. For higher order correlation functions entering the covariance computation, we rely on their respective one-halo component, which can be written in a general fashion as 
\begin{equation}
    P_{n}^{1\mathrm{h}}\left(\ell_1,\dots,\ell_{n}\right) = \int \diff z \ \mathcal{Q}^{(n)}(z)\ \text{I}_{n}^{0}\left(k^z_{\ell_1},\dots,k^z_{\ell_n};z\right).
\end{equation}

\section{Noise and foreground spectral shape}
\label{app:Foregrounds_spectra_shape}

Here we enumerate the energy and angular dependence of the spectral components we used in the forecasts for next generation surveys.
For convenience, we express the SED in terms of thermodynamic temperature rather than antenna temperature dividing them by the black-body derivative
\begin{equation}
    \mathcal{G}(x)=\frac{x^2 e^x}{(e^x-1)^2} \, .
\end{equation}
where $x=h\nu k_B^{-1} T_\text{CMB}^{-1}$.

\paragraph{Instrumental noise.}
    \begin{equation}
        \Theta^\text{N}(\nu)= \Delta T_\nu^2 \, ,
        \qquad
        \xi^\text{N} \rightarrow 0 \, , 
        \qquad
        C^\text{N}_\ell
        = 8 \ln 2 e^{\ell(\ell+1) \theta_\nu^2} \, \theta_\nu^2 \, ,
    \end{equation}
with $\theta_\nu = (1/ \sqrt{8 \ln 2} ) \, \pi/(180 \times 60) \, \text{FWHM}_\nu$,
where $\Delta T_\nu$ and $\text{FWHM}_\nu$ are the instrument thermodynamic-temperature-error and Full Width at Half Maximum (FWHM) measured in $T_\text{CMB}$ units and arcseconds respectively, for the channel $\nu$.

\paragraph{CMB.} 
    \begin{equation}
        \Theta^\text{CMB}(\nu)=1 \, ,
        \qquad
        \xi^\text{CMB} \rightarrow \infty \, , 
        \qquad
        C^\text{CMB}_\ell
        = C_\ell^{TT} \, ,
    \end{equation}
where $C_\ell^{TT}$ is the dimensionless CMB temperature power spectrum.
    
\paragraph{Free free.}
    \begin{equation}
    \begin{gathered}
        \Theta^\text{FF}(\nu) 
        \propto
        \left\{
            1 + \ln \left[1+ \left(\frac{\nu_\text{FF}}{\nu}\right)^{\sqrt{3}/\pi}\right]
        \right\}\mathcal{G}^{-1}(x) \, ,
        \quad
        \nu_\text{FF} = \SI{255.33}{GHz}\left(\frac{\SI{7000}{K}}{\SI{1000}{K}}\right)^{3/2} \, ,
    \\
        \xi^\text{FF} = 0.02 \, , 
        \qquad
        C^\text{FF}_\ell
        = \frac{(\SI{70}{\mu K})^2}{[\Theta^\text{FF}(\SI{31}{GHz})]^2} \ell^{-3} \, .
    \end{gathered}
    \end{equation}
    
\paragraph{Thermal dust.}
    \begin{equation}
    \begin{gathered}
        \Theta^\text{Dust}(\nu) 
        \propto
            x_\text D^{1.53} \frac{x^3}{e^x-1}
            \mathcal{G}^{-1}(x) \, ,
        \quad
        x_\text D = \frac{h\nu}{k_B \SI{21}{K}} \, ,
    \\
        \xi^\text{Dust} = 0.3 \, , 
        \qquad
        C^\text{Dust}_\ell
        = \frac{(\SI{24}{\mu K})^2}{[\Theta^\text{Dust}(\SI{90}{GHz})]^2} \ell^{-3} \, .
    \end{gathered}
    \end{equation}
    
\paragraph{Synchrotron.}
    \begin{equation}
    \begin{gathered}
        \Theta^\text{Synch}(\nu) 
        \propto
        \left(
            \frac{\nu}{\SI{100}{GHz}}
        \right)^{-0.82}
        \left[
            1+\frac{1}{2}0.2 \ln^2\left( \frac{\nu}{\SI{100}{GHz}} \right)
        \right]\mathcal{G}^{-1}(x)
    \\
        \xi^\text{Synch} = 0.15 \, , 
        \qquad
        C^\text{Synch}_\ell
        = \frac{(\SI{101}{\mu K})^2}{[\Theta^\text{Synch}(\SI{19}{GHz})]^2} \ell^{-2.4} \, .
    \end{gathered}
    \end{equation}
    
\paragraph{Radio point sources.}
    \begin{equation}
    \begin{gathered}
        \Theta^\text{Radio}(\nu) 
        \propto
        \nu^{-0.5}
        \mathcal{G}^{-1}(x) \, ,
    \\
        \xi^\text{Radio} = 0.5 \, , 
        \qquad
        C^\text{Radio}_\ell
        =
         \frac{(\sqrt{3}\ \mu \text{K})^2}{[\Theta^\text{Radio}(\SI{31}{GHz})]^2}
        \frac{2\pi}{\ell(\ell+1)} \frac{\ell^2}{3000^2} \, .
    \end{gathered}
    \end{equation}
    
\paragraph{Infrared point sources.}
    \begin{equation}
    \begin{gathered}
        \Theta^\text{IR}(\nu) 
        \propto
            x_\text{IR}^{0.86} \frac{x^3}{e^x-1}
            \mathcal{G}^{-1}(x) \, ,
        \quad
        x_\text{IR} = \frac{h\nu}{k_B \SI{18.8}{K}} \, ,
    \\
        \xi^\text{IR} = 0.3 \, , 
        \quad
        C^\text{IR}_\ell
        =
        \frac{2\pi}{\ell(\ell+1)}
         \frac{1}{[\Theta^\text{IR}(\SI{31}{GHz})]^2}
         \left[
         (\SI{7}{\mu K})^2\frac{\ell^2}{3000^2}
         +(\SI{5.7}{\mu K})^2\frac{\ell^{2-1.2}}{3000^{2-1.2}}
         \right] .
    \end{gathered}
    \end{equation}

\section{Full triangle plots}
\label{sec:A_FullTrianglePlot}
\begin{figure}
    \centering
    \includegraphics[width=\linewidth]{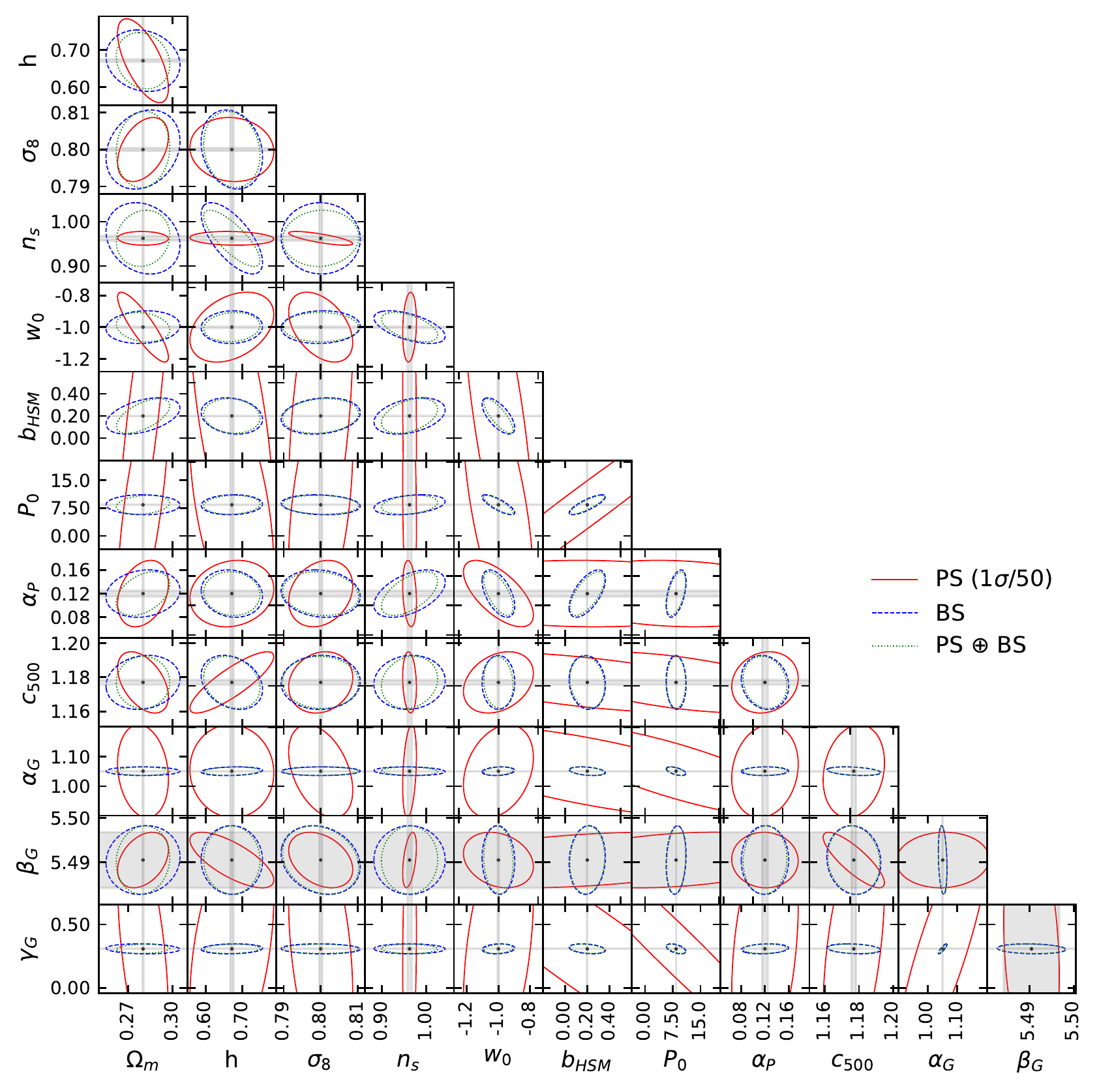}
    \caption{Triangle plots for a Cosmic variance limited experiment with perfect foreground separation, $\ell_\text{max}=5000$, and $f_\text{sky}=1$. Notice that the power spectrum $1\sigma$ ellipses have been rescaled by a factor to fit in the same graph.
    The grey bands are the power spectrum conditional errors.
    Figure \ref{fig:NoNoiseNoForeground5000} is a sub-sample of this one.
    }
    \label{fig:NoNoiseNoForeground5000_all}
\end{figure}

\begin{figure}
    \centering
    \includegraphics[width=\linewidth]{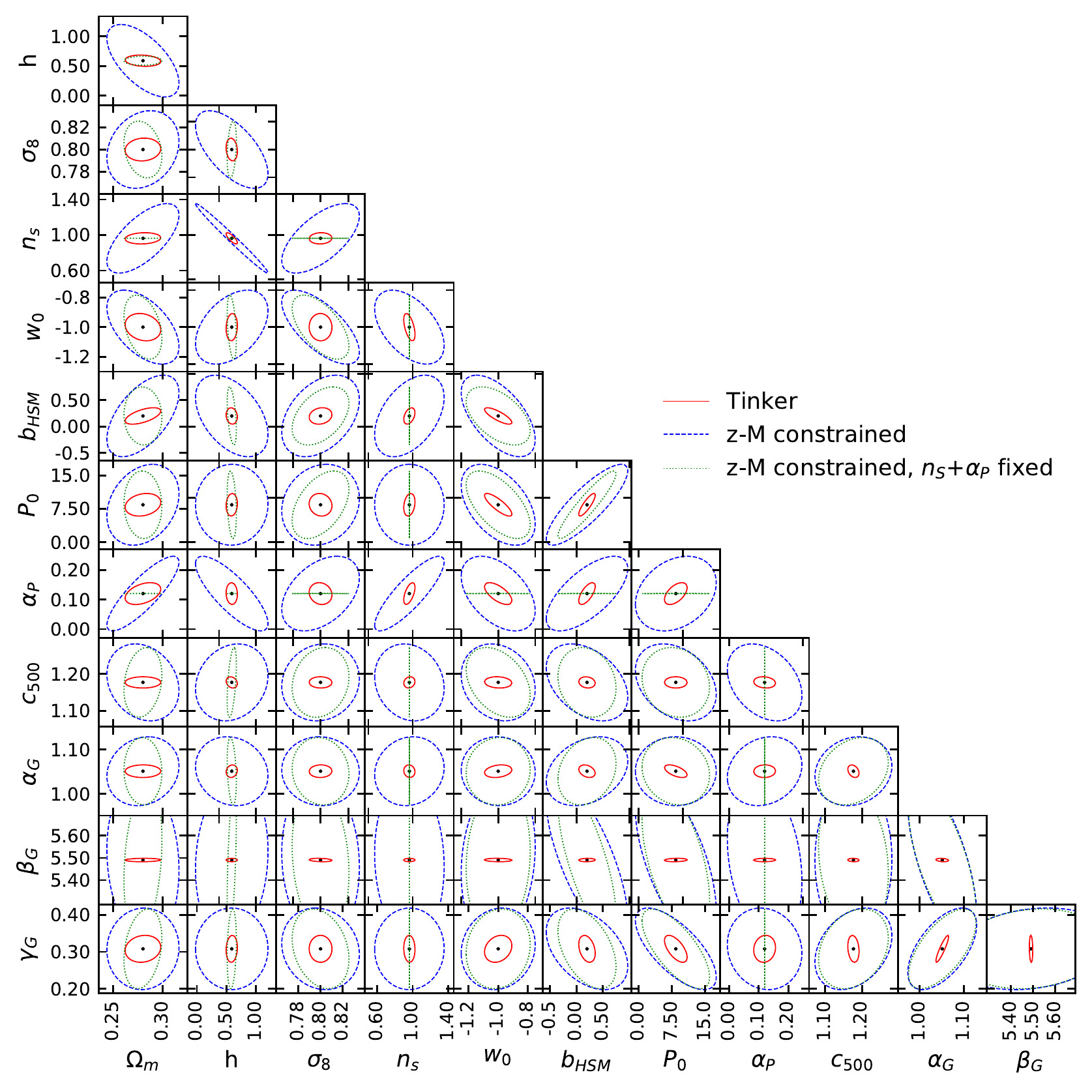}
    \caption{Triangle plot for the joint power spectrum and bispectrum analysis with a CVL survey with $\ell_\text{max}=5000$ and $f_\text{sky}=1$. The figure propose a comparison of the two models detailed in section \ref{sec:Results:CVL}.
    We also try to fix $n_s$ and $\alpha_P$ to show the impact of the most degenerate parameters, that could be determined using external data.}
    \label{fig:NoNoiseNoForeground5000_Tinker_zMCut}
\end{figure}

For completeness we report here the triangle plot for all the parameters.
In figure \ref{fig:NoNoiseNoForeground5000_all} we compare the power spectrum and bispectrum covariance ellipses for a noiseless survey with $\ell_\text{max}=5000$, $f_\text{sky}=1$, and no foreground contamination, assuming the full integration domain in redshift and masses from \cite{2010ApJ...724..878T}.
A comparison to the case in which the integration boundaries have been set according to the \Planck observation is drawn in figure \ref{fig:NoNoiseNoForeground5000_Tinker_zMCut}.

\begin{figure}
    \centering
    \includegraphics[width=\linewidth]{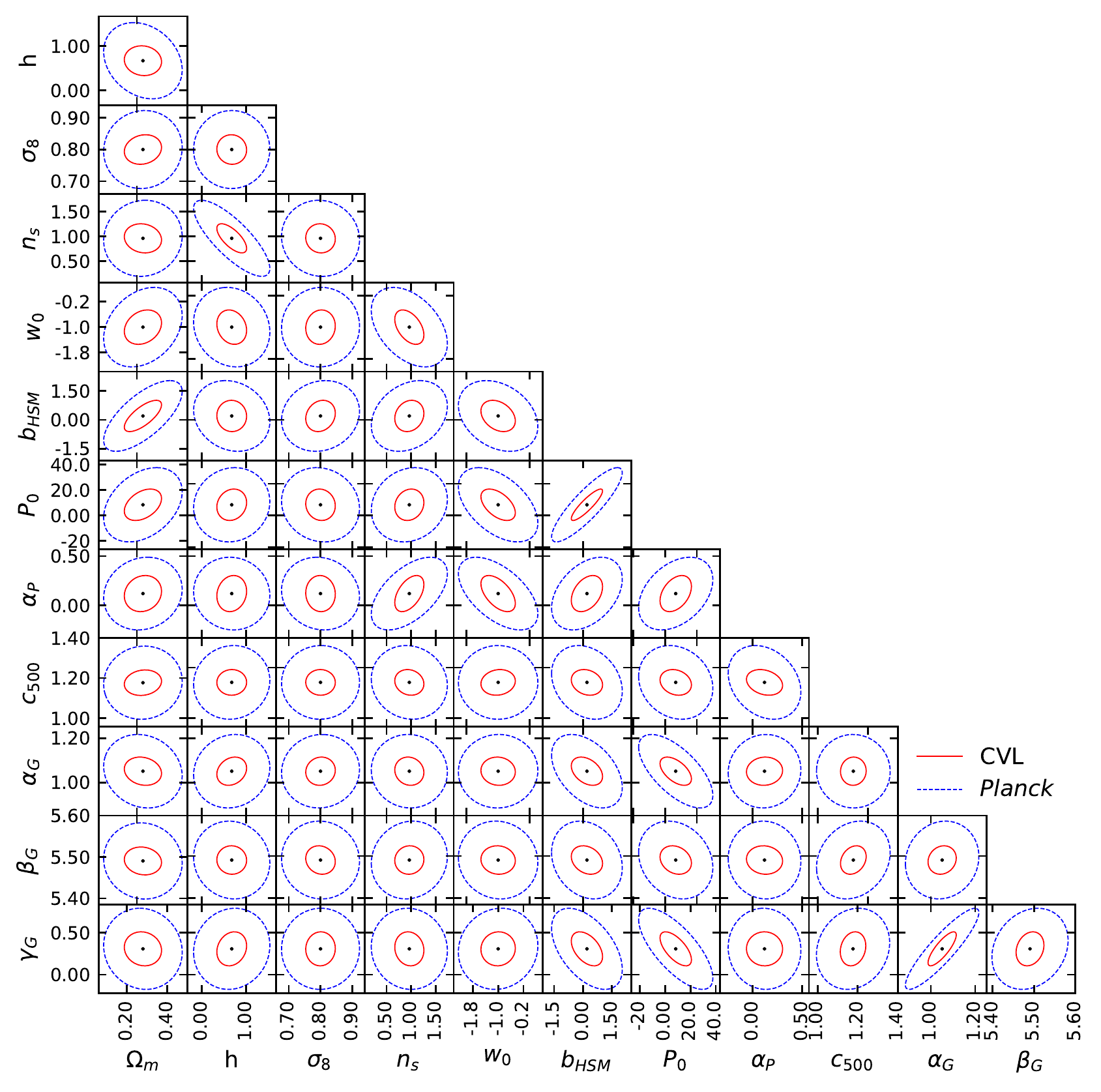}
    \caption{Triangle plot for the joint power spectrum and bispectrum analysis with a CVL survey with $\ell_\text{max}=1000$ and $f_\text{sky}=1$, compared with \textit{Planck}.}
    \label{fig:CVL1000_vs_Planck}
\end{figure}

\begin{figure}
    \centering
    \includegraphics[width=\linewidth]{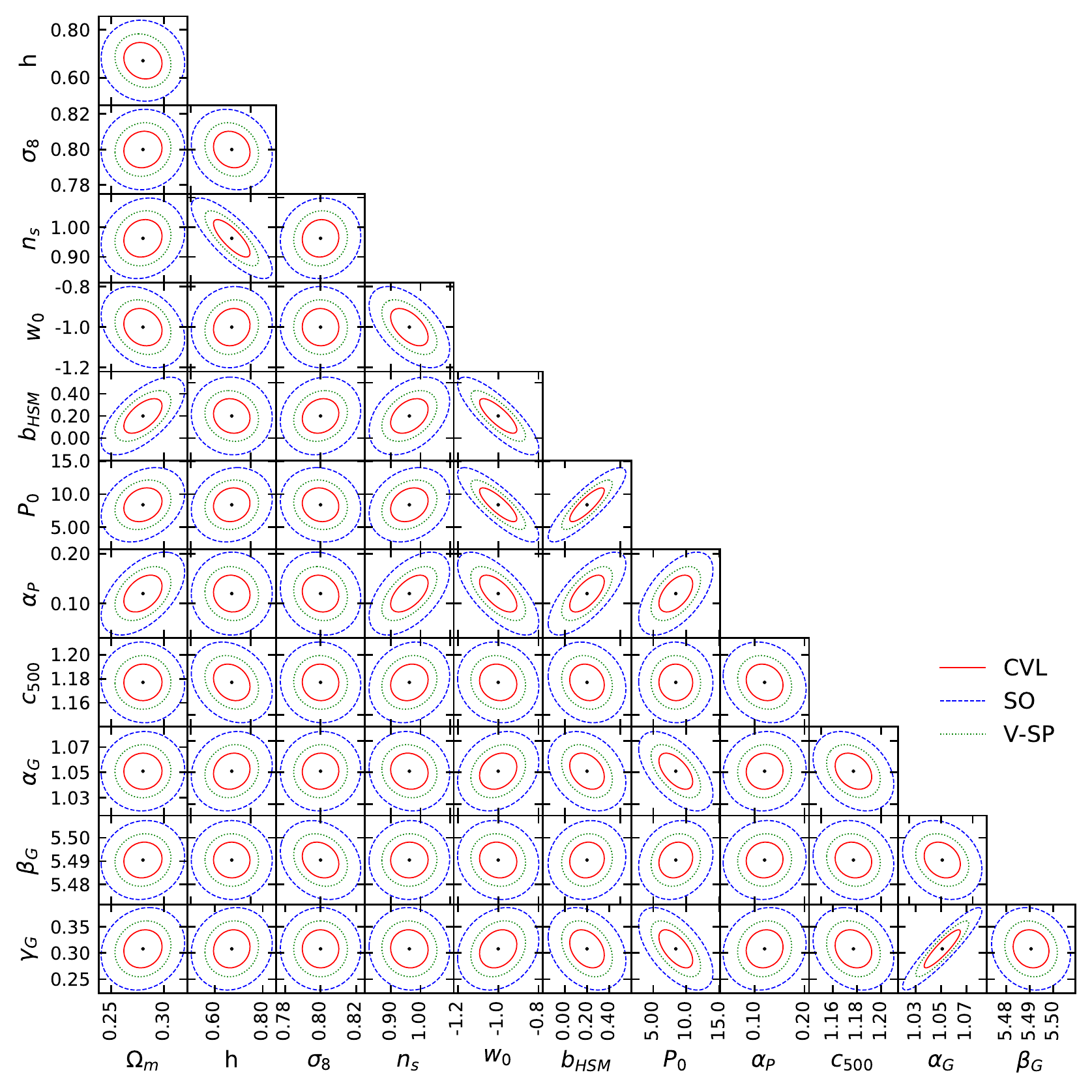}
    \caption{Triangle plot for the joint power spectrum and bispectrum analysis with a CVL survey with $\ell_\text{max}=5000$, compared with SO and V-SP.
    they have $f_\text{sky}=1$, 0.30, and 0.47,  respectively.}
    \label{fig:CVL5000_vs_SO_vs_VSP}
\end{figure}

Figure \ref{fig:CVL1000_vs_Planck} and figure \ref{fig:CVL5000_vs_SO_vs_VSP} show the triangle plots for actual surveys, taking into account both instrumental noise and foreground contamination; \Planck in figure \ref{fig:CVL1000_vs_Planck}, and SO and V-SP in \ref{fig:CVL5000_vs_SO_vs_VSP}.
For reference, we also show there the results for the CVL survey with $\ell_\text{max}=1000$ and 5000 respectively.

\section{Binning}
\label{A_Binning}

As there is not a general recipe to choose a priori the best binning scheme, we validated our choice both calculating the correlation between the binned and unbinned bispectrum, and by comparing different schemes.

In general, one can evaluate the correlation between two signals, and verify if they can be distinguished from each other using a given set of data, calculating theirs scalar product using the covariance matrix as metric
\cite{Smith:2006ud, Fergusson:2008ra, Fergusson:2009nv}
\begin{equation}
    \text{Corr}(b,b')
    \equiv
    \frac{1}{\mathcal{N}} \,
    \vec{b} \cdot \text{Cov}[b,b] \vec{b}',
    \qquad
    \mathcal{N}=
    \sqrt{\vec{b} \cdot \text{Cov}[b,b] \vec{b} \,} \,  
    \sqrt{\vec{b}' \cdot \text{Cov}[b,b] \vec{b}' \,} \, .
\end{equation}
In \cite{Smith:2006ud, Fergusson:2008ra, Fergusson:2009nv}, this method was employed in the case of the CMB primary bispectrum. That relatively simpler case allowed them to carry an exact calculation of the correlation.
In our case this is not possible, and we have to resort to two approximations.
First, in principle, the covariance would not depend on the tested models.
In the case of the primary bispectrum, the weakly non-Gaussian limit can be employed, so that the covariance only depends on the $C_\ell$ and instrumental noise, both of which are independent of the bispectrum.
For an actual survey, the covariance could instead be constructed by the collected data and, as such, it would again be independent from the theoretical model.
Instead, in our case, we assume that the covariance is the one calculated using our fiducial model, and we do not vary it while we compare another model to the fiducial one.
Second, the theoretically sound way to test the binning would be to compare the binned bispectrum to the full bispectrum calculated on every multipole, just repeating the representative value of each bin for every configuration in the bin.
However, this leads to the problem that the covariance matrix quickly becomes ill-conditioned when the binning is made finer and finer.
This problem was again not present in the case of the primary bispectrum since in the weakly non-Gaussian limit the covariance matrix is diagonal and, as such, trivial to invert.
To overcome this problem, we exploit the fact that the bispectrum is a monotonous function of $(\ell_1, \ell_2, \ell_3)$.
The values of the bispectrum in the bin that deviate the most from the bin representative will therefore be the two calculated at the extreme bin boundaries, e.g. (100,100,100) and (150,150,150) for the bin $[100,150]\times[100,150]\times[100,150]$.
Therefore, we compare the fiducial bispectrum $b$ with the bispectrum calculated on the higher boundary $b^\text{hi}$ and the lower one $b^\text{low}$.
We obtain
\begin{equation}
    \text{Corr}(b,b^\text{hi}) = 99\% \, ,
    \qquad
    \text{Corr}(b,b^\text{low}) = 97\% \, .
\end{equation}

We have verified that changing the maximum and minimum multipoles ($[10, 5000]$, $[70, 5000]$, $[10, 3000]$), the number of bins (27, 25, 32) has little effect ($\lesssim 10\%$) on the signal to noise ratio in the case of logarithmically spaced bins. 
The same applies when using (according to \cite{Lacasa:2014gha}) a combination of linearly and logarithmically spaced bins (linear with $\Delta \ell = 64$ in $[32, 1440]$, 10 logarithmically spaced ones in $[1440, 5000]$).

\section{Bispectrum derivatives and breaking the degeneracies}
\label{sec:A_BispectrumKernel}

\begin{figure}
    \centering
    \includegraphics[width=\linewidth]{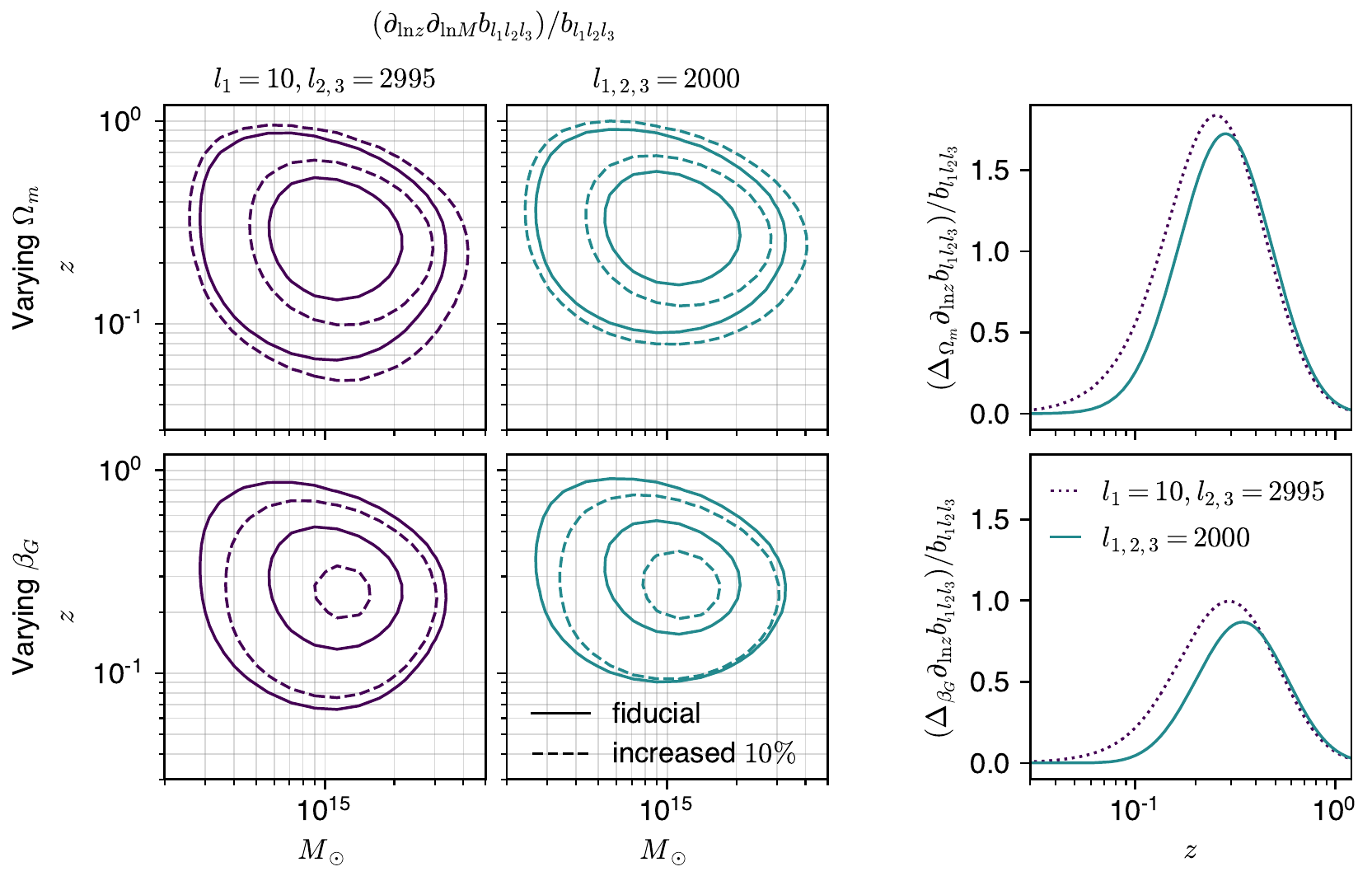}
    \caption{Effect of parameter variations on the bispectrum kernel for two isoperimetric configurations: (10,2995,2995) and (2000,2000,2000).
    In the top row we assess the change due to an increase in $\Omega_m$, and in the bottom row an increase in $\beta_G$.
    \emph{Left and centre}) contour plot of the bispectrum kernel in the redshift-mass plane.
    Solid lines refer to the kernel calculated with the fiducial set of parameters, dashed lines to the kernel calculated with $\Omega_m$ or $\beta_G$ increased by 10\%.
    The two left panels display a squeezed bispectrum configuration, while the two central panels an equilateral one.
    The contour lines are traced at 1/10th and half maximum.
    \emph{Right}) finite differences, as defined in eq. \eqref{eq:finite_differences}, of the kernel integrated over all masses.
    }
    \label{fig:zMDerivative}
\end{figure}

To partially motivate the difference among the derivatives of squeezed and equilateral bispectrum, we investigated the impact of parameter changes on the bispectrum kernel.
Not to have to deal with numerical convergence of derivatives in each point we consider the bispectrum kernel evaluated with the fiducial parameters, compared with the same evaluated with one parameter increased by 10\%, and  finite differences defined as
\begin{equation}
    \Delta_\theta \frac{\partial b_{\ell_1 \ell_2 \ell_3}}{\partial \ln z}
    =
    \frac{\partial b_{\ell_1 \ell_2 \ell_3}}{\partial \ln z } (\theta, \tilde{\theta})
    -
    \frac{\partial b_{\ell_1 \ell_2 \ell_3}}{\partial \ln z} (1.1 \theta, \tilde{\theta})
\label{eq:finite_differences}
\end{equation}
where $\theta$ is the parameter being varied and $\tilde{\theta}$ is the vector of all the others which have been kept fixed.
We already argued that isoperimetric bispectrum configurations probe quite similar regions in the $z$-$M$ plane, but here we need to focus on the subtle differences that are still present.
In the left four panels of figure \ref{fig:zMDerivative} we therefore compare the bispectrum kernel of the isoperimetric (10, 2995, 2995) and (2000, 2000, 2000) configurations, calculated for different combination of parameters.
Different parameters modify the kernel in different way (first vs second row), but also different configurations react differently (left vs central column).
To see this, we show contour lines for the kernel calculated with the parameter fiducial value (solid) and with one of the parameters ($\Omega_m$ on top, $\beta_G$ at the bottom) increased by 10\%.
This can be better appreciated in the right column, where we carried out the mass integral and plotted the finite differences as defined in eq. \eqref{eq:finite_differences}.
The bispectrum derivatives (which are in a sense related to the integral of the curves plotted there) are not only different between configurations (squeezed or equilateral) but different parameters affect the various configurations differently (the ratio of the curves in the top panel is different from the ratio of the two curves in the bottom panel).

\bibliographystyle{JHEP}

\bibliography{bibliografia}

\end{document}